\documentclass[draft]{agujournal2018}
\usepackage{apacite}
\usepackage{lineno}

\usepackage{graphicx}
\usepackage{color}
\usepackage{multirow}
\usepackage{amsmath,amsfonts,amssymb,mathtools}

\draftfalse

\journalname{JGR:Oceans}

\newcommand{\bU}{\mathbf{U}}

\newcommand{\bk}{\mathbf{k}}

\newcommand{\ctil}{\tilde{\mathbf{c}}}

\newcommand{\ci}{\tilde{c}_{i}}

\newcommand{\rev}[1]{#1}
\newcommand{\dl}[1]{}
\newcommand{\drev}[2]{#2}

\newcommand{\rv}[1]{{#1}}
\newcommand{\del}[1]{}
\newcommand{\drv}[2]{#2}

\newcommand{\be}{\begin{equation}}
\newcommand{\ee}{\end{equation}}

\begin{document}

%
%


\title{
An improved method for determining near-surface currents from wave dispersion measurements}

%
%




\authors{B. K. Smeltzer\affil{1}, E. \AE s\o y\affil{1}, Anna \AA dn\o y\affil{1}, S. \AA. Ellingsen\affil{1}}


\affiliation{1}{Department of Energy and Process Engineering, Norwegian University of Science and Technology, Trondheim, Norway}




\correspondingauthor{Benjamin K. Smeltzer}{benjamin.smeltzer@ntnu.no}




\begin{keypoints}
\item Simple and more accurate method for reconstructing near-surface current profiles from wave spectra
\item Laboratory setup where shear currents and waves can be well-controlled and measured
\item Novel adaptation of a method for extracting wavelength-dependent Doppler shifts from wave spectra
\end{keypoints}

%
%


\begin{abstract}
A new inversion method for determining near-surface shear currents from a measured wave spectrum is introduced. The method is straightforward to implement and starts from the existing state-of-the-art technique of assigning effective depths to measured wavenumber-dependent Doppler shift velocities. A polynomial fit is performed, with the coefficients scaled based on a simple derived relation to produce a current profile that is an improved estimate of the true profile. The method involves no user-input parameters, with the optimal parameters involved in the polynomial fit being chosen based on a simple criterion involving the measured Doppler shift data only. The method is tested on experimental data obtained from a laboratory where current profiles of variable depth dependence could be created and measured by particle image velocimetry, which served as ``truth'' measurements. Applying the new inversion method to experimentally measured Doppler shifts resulted in a $>3\times$ improvement in accuracy relative to the state-of-the-art for current profiles with significant near-surface curvature. The experiments are dynamically similar to typical oceanographic flows such as wind-drift profiles and our laboratory thus makes a suitable and eminently useful scale model of the real-life setting. Our results show that the new method can achieve improved accuracy in reconstructing near-surface shear profiles from wave measurements by a simple extension of methods which are currently in use, incurring little extra complexity and effort. A novel adaptation of the normalized scalar product method has been implemented, able to extract Doppler shift velocities as a function of wavenumber from the measured wave spectrum.

\end{abstract}

%
%

%


%
%
%
%

\section{Introduction}

Characterizing near-surface ocean currents is of importance to a vast range of applications. At a fundamental scientific level, near-surface currents influence the exchange of energy and momentum between the air and sea \citep{terray96, kudryavtsev08}, impacting climate models. At a more practical level, currents affect wave-body forces, and can be relevant for operational safety in coastal areas \cite{dalrymple73,zippel17}. Accurate measurements of the mean flow in the top meters of the water column are difficult to obtain, in large part due to the presence of waves which induce platform motions and additional sources of noise. Conventional methods such as acoustic Doppler current profiling (ADCP) typically discard data in the topmost few meters of the water column.

An attractive alternative to in situ techniques is to deduce currents from measurements of waves, whose dispersion is altered by the presence of a background flow. The approach has the advantage of enabling remote sensing methods such as radar or optical-based detection, with the potential for mapping currents over a larger area (multiple $\mathrm{km}^2$) compared with point measurements. In addition, waves are most sensitive to currents near the free surface, precisely the regime where other conventional methods such as ADCP struggle. The vast majority of wave-based near-surface current measurements reported in the literature have used radar, including high frequency (HF) radar \citep[e.g.,][]{crombie55, young85, stewart74, ha79, fernandez96,teague01,shrira01} and more recently X-band radar systems \citep [e.g.,][]{young85,gangeskar02,lund15,campana16,campana17,lund18}, also in some cases to reconstruct the bathymetry \citep[e.g.,][]{hessner14,hessner09}. Optical methods have also been used to a lesser extent \citep{dugan01,dugan03,laxague17,horstmann17,laxague18}.

Though wave-based current measurements offer several distinct advantages compared to other methods, they have a number of inherent challenges. Firstly, determining the current profile as a function of depth without stringent \emph{a priori} assumptions as to the functional form requires the ability to measure waves over a spectrum of wavelengths and directions. The quality of results is thus dependent on the sea state \citep{lund15,campana16,campana17}. Secondly and more fundamentally, the determination of the current depth profile from wave dispersion measurements is a mathematically ill-posed inverse problem. The inferred current profile is not mathematically unique, and noise in wave measurements gets amplified in the inversion process \citep{ha79}. As a result, a priori assumptions and constraints of the depth-dependence of the current profile based on physical intuition have typically been imposed. 

Despite \drv{the hardships}{these difficulties}, wave-based current measurements have been used in the field for many decades. The techniques involve reconstructing the near-surface current from measured alterations to the wave frequency, and are termed ``inversion methods." The most common and elementary methods involve determining a single current vector representative of a weighted average of the near surface flow, with other more recent methods reconstructing some degree of detail as to the depth-dependence of the flow. In reviewing the previously developed inversion methods, we first consider the dispersion relation for small-amplitude linear waves propagating atop a depth-varying flow, which can be approximated as:
\begin{equation}
\omega_{\mathrm{DR}}(\mathbf{k}) = \omega_0(k) + \bk\cdot\ctil(k),
\label{eq:dr}
\end{equation}
where $\omega_{\mathrm{DR}}$ is the wave frequency, $\omega_0$ the frequency in quiescent waters, $\bk = \{k_x,k_y\}$  the wavevector, $k = |\bk|$, and $\ctil$ a wavenumber-dependent Doppler shift velocity due to the background current. \drv{Presume}{The} $z=0$ \rv{plane} is the undisturbed water surface and the bottom is found at $z=-h$ with $h>0$. We shall mostly work in the deep water regime $kh \gtrsim \pi$ where $h = \infty$ can be assumed. As first shown by \citet{stewart74}, the Doppler shift can be approximated as a weighted average of the current profile as a function of depth as
\begin{equation}
\ctil(k) = 2k\int_{-\infty}^0\bU(z)\mathrm{e}^{2kz}dz,
\label{eq:SJ}
\end{equation}
where $\bU(z) = [U(z),V(z)]$ is the current profile. The finite depth version of the Stewart \& Joy (SJ) approximation \eqref{eq:SJ} was derived by \citet{skop87} and extended by \citet{kirby89}. The weighting term decays exponentially with depth (in deep water), reaching a value of 0.2\% \drv{that at}{of} the surface \rv{value} at a depth equal to half the wavelength ($kz = -\pi$). Short wavelengths are thus sensitive only to currents near the surface, whereas longer wavelengths are affected by currents at greater depths. The inversion method involves using values of $\ctil(k)$ obtained from experimental data to determine the unknown $\bU(z)$. 

A word of warning is warranted when referring to $\ctil$ as the ``Doppler shift'' as is conventional. While $\ctil$ occurs in \eqref{eq:dr} exactly as would a Doppler frequency shift resulting from Galileian transformation upon changing reference system, it should not be interpreted as such. A misunderstanding has arisen from this name that the same Doppler shift should also be added to the wave's group velocity to account for the shear, but this is not correct as pointed out by \cite{banihashemi17}. Rather, the group velocity remains $d \omega/d k$, for which taking the $k$-dependence of $\ctil$ into account is key. 
We shall follow the numenclatorial convention in the literature and refer to $\ctil$ as the Doppler shift velocity while bearing this in mind.

Various wave detection methods are sensitive to different spectral ranges of $k$ and have led to the development of a number of inversion methods. In the case of HF radar, the detected signal is dominated by resonant Bragg scattering, effectively measuring the Doppler velocity of a surface wave with a wavelength half that of the radar system. Data reported from single-frequency HF radar is often referred to as the surface current, yet \del{is }more precisely \rv{it is} a weighted average of the current profile from (\ref{eq:SJ}), as it essentially measures $\ctil(k_{\mathrm{HF}})$ ($k_{\mathrm{HF}}$ being the wavenumber of the resonant wave) without information concerning the depth-dependence. Depth-profile information can be obtained by using multiple radar frequencies \citep{stewart74,ha79,fernandez96,teague01} which probe different resonant wavenumbers. Similarly, other detection methods such as X-band radar or optical techniques inherently measure a wide spectrum of wavelengths, thus evaluating (\ref{eq:SJ}) at many $k$-values and enabling the use of inversion methods to estimate the current depth-dependence. 

Inversion methods of determining $\bU(z)$ from a set of measured values of $\ctil_i = \{\tilde{c}_{x,i},\tilde{c}_{y,i}\}$ at discrete wavenumbers $k_i$ can be carried out separately for each velocity component, i.e. $U(z)$ can be found from $\tilde{c}_{x,i}$, and $V(z)$ from $\tilde{c}_{y,i}$. To ease the notation, in the following we outline the new inversion method using $U(z)$ and $\ci$ to denote the flow velocity and Doppler shift velocities, with the implicit understanding that they may correspond to either dimension in the horizontal plane. The subscript $i$ indicates that the respective variable takes on a discrete set of values as may be extracted from experimental data.

Assuming a given functional form to the current profile, one can assign effective depths to the measured Doppler velocities based on the wavenumber \rv{by finding the depth at which the Doppler velocity is equal to the current}, i.e. $\ci = U\rv{(}Z_{\mathrm{eff}}(k_i)\rv{)}$. For the commonly assumed case of a current profile which varies linearly with depth, $U(z) = Sz + U_0$\rev{, where $S$ is the vorticity and $U_0$ the surface current}. By the approximation \eqref{eq:SJ} the Doppler shift is approximated as

\begin{equation}\label{EDM}
\ci = -\frac{S}{2k_i} + U_0 = \rv{U\left(z = -(2k_i)^{-1}\right)}. 
\end{equation}
The last form shows that assuming linear current, deep water and using the SJ approximation, the appropriate effective depth is  
\be\label{zlin}
  Z_{\mathrm{eff, lin}}(k) = -(2k)^{-1}. 
\ee
(In other words $Z_{\mathrm{eff}}(k)$ is approximately $8\%$ of the wavelength.) A similar relation can also be derived for a logarithmic profile \citep{plant80}. \drv{Estimating}{We refer to the method of estimating} $U(z)$ from a measured $\ci(k)$ using \eqref{EDM} or its sibling assuming a logarithmic profile \del{we refer to }as the effective depth method (EDM). The EDM has been used extensively in the literature for estimating near-surface shear currents \citep[e.g.,][]{teague01,lund15,laxague17,laxague18,stewart74,fernandez96}.

A clear weakness of the EDM, however, is that it relies on assumptions as to the functional form of the depth dependence. \citet{ha79} developed a method for inverting (\ref{eq:SJ}) directly based on a series of measured $\ctil$ values, which was further developed and applied to data from X-band radar by \citet{campana16}. The method involves a Legendre quadrature approximation to the integral, with constraints on the curvature of the current profile as well as the distance from an initial guess in order to suppress the amplification of experimental noise. The method avoids initial assumptions as to form of the current profile and yields current estimates at greater depths. 
The reconstructed $U(z)$ has comparable accuracy relative to the EDM when compared against ADCP ``truth" measurements.

We present a new inversion method which is completely free of parameters. The method, which is derived assuming deep water, uses the current profile obtained by the EDM, and fits it to a polynomial function\drv{,}{. The method} then makes use of a simple relation which follows from \eqref{eq:SJ} to construct an improved estimate of the true profile $U(z)$ directly from the coefficients of the fit. The method is validated and tested on experimental data from a laboratory setup, where the background current velocity profile and wave spectrum could be well-controlled and characterized.

In the following we describe the new method in Section \ref{sec:pedm}, and examine its performance also in finite water depth. Section \ref{sec:exp} describes the experimental setup and analysis of the data, where an adapted version of a normalized scalar product (NSP) method is used to extract Doppler shifts from wave spectra. Section \ref{sec:expres} demonstrates \rv{the} use of the new inversion method on experimentally measured Doppler shifts, where in situ measurements of the current profile are used as ``truth'' measurements \drv{to compare against}{for validation}. The performance of the method is evaluated by considering the fractional decrease in error of the depth profile achieved by the new inversion method compared to the EDM.


\section{Polynomial effective depth method}
\label{sec:pedm}


From experimental data of the wave spectrum, a set of Doppler shift velocities $\ctil_i$ at unique wavevector magnitudes $k_i$ can be obtained by a number of methods such as least squares techniques \citep{campana17,senet01}, or a normalized scalar product (NSP) method \rev{\citep{huang16,huang12,serafino10}} used herein (described in section \ref{sec:exp}).

Assuming a polynomial current profile of the form $U(z) = \sum_{n = 0}^\infty u_n z^n$ in deep water, evaluation of (\ref{eq:SJ}) yields the SJ approximation
\begin{equation}
  \tilde{c}(k) = \sum_{n = 0}^\infty n!u_n \left(-\frac{1}{2k}\right)^n
\label{eq:pedm_ctil}
\end{equation}
for the Doppler shift velocities. We notice that $(-2k)^{-1}$ is equal to the mapping function $Z_{\mathrm{eff, lin}}(k)$ used in the EDM assuming a linear frofile, equation \eqref{zlin}. Using the EDM with this mapping the estimated current profile is
%
%
%
\begin{equation}
  U_{\mathrm{EDM}}(z) = \sum_{n = 0}^\infty n!u_n z^n.
\label{eq:pedm}
\end{equation}
Thus, the mapped profile $U_{\mathrm{EDM}}(z)$ is also of polynomial form with coefficients of the $n$-th order term differing by a factor $n!$ from those of the true profile $U(z)$. The estimated velocity profile $U_{\mathrm{EDM}}(z)$ will suffer from inaccuracies since the mapping function is not the correct one. The new inversion method, referred to hereafter as the polynomial effective depth method (PEDM), seeks to improve this by making use of the simple relationship between the coefficients in the series representation of $U_{\mathrm{\del{P}EDM}}(z)$ and the true profile $U(z)$, namely that they differ by a factor $n!$.

Explicitly, the PEDM procedure consists of the following three steps:
\begin{enumerate}
  \item \drev{From the measured values of $\ci(k)$ at wavenumbers $k_i$, obtain $U_{\mathrm{EDM}}(z)$ using \eqref{EDM}}{For each of the measured values $\ci$, assign effective depths $z_i = -(2k_i)^{-1}$ according to the EDM procedure of (\ref{EDM}) and (\ref{zlin})}.
  \item \drev{Fit the obtained values of $U_{\mathrm{EDM}}(z)$}{Obtain $U_{\mathrm{EDM}}(z)$ by fitting the set of points $\{z_i,\ci\}$} to a polynomial of degree $n_\mathrm{max}$:
    \be\label{EDMfit}
      U_{\mathrm{EDM}}(z)\approx\sum_{n=0}^{n_\mathrm{max}} u_{\text{EDM},n}z^n\drv{.}{,}
    \ee
    \rv{where $u_{\text{EDM},n}$ are the coefficients obtained in the polynomial fit.}
  \item Then the improved PEDM estimate is 
    \be\label{PEDM2}
      U_{\mathrm{PEDM}}(z)=\sum_{n=0}^{n_\mathrm{max}} \frac1{n!} u_{\text{EDM},n}z^n.
    \ee
    
\end{enumerate}
\rv{Equation (\ref{PEDM2}) follows immediately from a comparison of (\ref{eq:pedm}) and (\ref{EDMfit}), where $u_{\text{EDM},n} = n!u_n$.}

\subsection{Theoretical limitations}

Two notable potential complications arise: finite depth where (\ref{eq:pedm_ctil}) and (\ref{eq:pedm}) are no longer strictly valid, and realistic situations where errors in experimentally measured Doppler shifts, which in addition are measured at a finite range of wavenumbers, lead to errors in the fitted polynomial coefficients rapidly increasing for higher values of $n$.

\begin{figure}[htb]
\centering
	\includegraphics{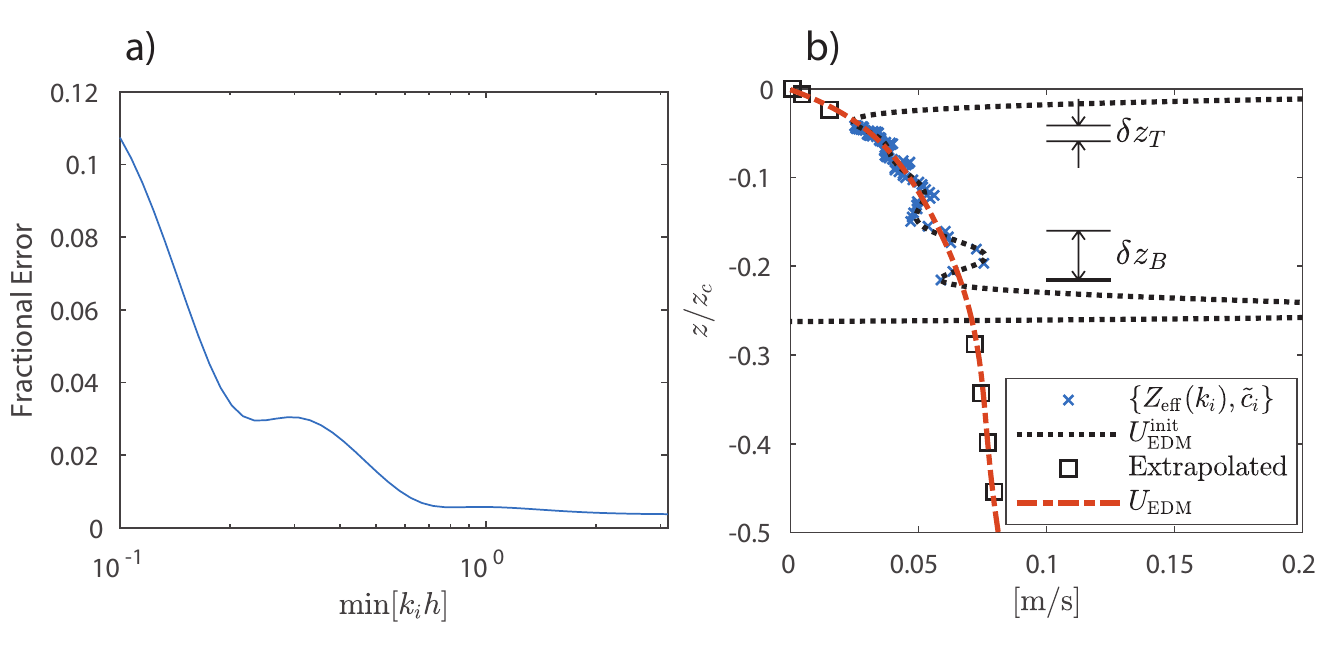}
	\caption{\rev{a)} Relative root mean square (RMS) error between the PEDM and true profiles of exponential form as a function of water depth normalized to the minimum mapped wavenumber. \rev{b) Illustration of the parameters involved in practical implementation of the PEDM, as part of the 6-step process described in section \ref{sec:pedm_impl}.}}
	\label{fig:pedm_fd}
\end{figure}

\subsubsection{Performance for finite depth}

In the case of finite depth $h$, an explicit relation of the form of (\ref{eq:pedm}) cannot be derived since the mapping function $Z_{\mathrm{eff}}(k) = -(2k)^{-1}\tanh kh$ in finite depth cannot be solved with respect to $k$ analytically\rev{, but must be inverted numerically. The approximation (\ref{eq:SJ}), moreover, obtains a more complicated form less amenable to analytical treatment \citep{skop87}}. To examine the effect of finite depth on the accuracy of the PEDM, we consider an exponential profile of the form $U(z) = U_0\exp(\alpha z)$, with $\alpha$\drev{ equal to a value twenty times that of the minimum wavenumber $\mathrm{min}[k_i]$}{$=8\cdot\mathrm{min}[k_i]/\tanh( \mathrm{min}[k_ih]) $ to preserve the same functional form within the range of mapped depths regardless of the water depth}, and $U_0$ the surface current. 

We consider the implementation of the PEDM in \drev{infinite}{finite} depth with $n_{\mathrm{max}}=10$, simply using \drev{(\ref{PEDM2}) with $U_{\mathrm{EDM}}$ determined using the mapping function in deep water}{the finite depth mapping function in step 1 of section \ref{sec:pedm} to assign effective depths $z_i = -(2k_i)^{-1}\tanh k_ih$. Steps 2-3 of the PEDM procedure were unchanged}. The fractional depth-integrated root mean square (RMS) error between $U_{\mathrm{PEDM}}(z)$ and $U(z)$ was calculated for cases over a range of water depth values $\mathrm{min}[k_i]h$, with the results shown in Figure~\ref{fig:pedm_fd}\rev{a}. For all but the shallowest depths considered here, the deep water mapping function results in errors at the 1\% level. \dl{It is noted that for values of $\mathrm{min}[k_i]h<0.5$, the deep water mapping case results in depths $Z_{\mathrm{eff}}(k)$ being mapped below the bottom for some $k$; these points were discarded. }For most \dl{all }realistic combinations of water depth and relevant wavenumbers, Figure~\ref{fig:pedm_fd}\rev{a} indicates that the \drev{deep water}{finite depth} mapping function and (\ref{eq:pedm}) yield sufficient accuracy.

\subsubsection{Effect of limitations of measured Doppler shifts}
\label{sec:pedm_impl}

As mentioned, the fact that $\ci(k)$ is measured for a finite range of wavenumbers will affect accuracy. This is true of any inversion method for reconstructing $U(z)$ from dispersion measurements. 

To handle the realistic case of experimentally measured Doppler shifts at a finite range of wavenumbers, we \drev{perform}{extend the three-step process described in section~\ref{sec:pedm} to} a \drev{three}{6-}step process \rev{(the first three steps of which are illustrated schematically in Figure~\ref{fig:pedm_fd}b)}: \dl{first fitting the mapped Doppler shifts to a polynomial of order $n_{\mathrm{max}}$, then creating additional velocity-depth pairs by extrapolating up to the surface and down to a cutoff depth $z_c$, and finally by performing a second polynomial fit on the expanded set of velocity-depth points.}

\rev{
\begin{enumerate}
\item Fit the mapped Doppler shifts to a polynomial of order $n_{\mathrm{max}}$ to produce the profile $U_\mathrm{EDM}^\mathrm{init}(z)$ (Steps 1-2 in section \ref{sec:pedm}), using the finite depth mapping function if appropriate.
\item Create additional velocity-depth pairs by linearly extrapolating up to the surface and down to cutoff depth $z_c$. The extrapolation is performed based on \drv{the average shear of}{a linear fit to} $U_\mathrm{EDM}^\mathrm{init}(z)$ over a depth interval $\delta z_T$ and $\delta z_B$ at the shallow and deep end of the regime of mapped depths respectively, denoted in Figure~\ref{fig:pedm_fd}b. The extrapolated points are shown as the black squares.
\item Perform a second polynomial fit on the expanded set of points (also of order $n_{\mathrm{max}}$) to produce the profile considered to be $U_\mathrm{EDM}$.
\item Scale polynomial coefficients defining $U_\mathrm{EDM}$ by $n!$ as in (\ref{PEDM2}) to produce a profile $U_\mathrm{PEDM}^\mathrm{init}(z)$.
\item Create a new set of linearly extrapolated points down to $z_c$ based on \drv{the average shear of}{a linear fit to} $U_\mathrm{PEDM}^\mathrm{init}(z)$ \drv{in}{over} a depth region $\delta z_B/2$ at the deep end of the regime of mapped depths. Extrapolation is not performed up to the surface (thus differing from step 2).
\item Perform a final polynomial fit on the set of points including $U_\mathrm{PEDM}^\mathrm{init}(Z_\mathrm{eff}(k_i))$ and the extrapolated points in Step 5, to produce $U_\mathrm{PEDM}$.

\end{enumerate}
}

\dl{The extrapolation is performed based on the average shear of the initial polynomial fit current profile over a depth interval $\delta z_T$ and $\delta z_B$ at the shallow and deep end of the regime of mapped depths respectively. A set of velocities are calculated at intervals $\delta z_T$ and $\delta z_B$ up to $z = 0$ and down $z = -z_c$ respectively. The resulting polynomial coefficients from the second polynomial fit on the expanded set of points are considered the profile $U_\mathrm{EDM}$, and are then scaled by $n!$, whereby the above process is repeated with extrapolation only being performed at the deep end of the regime using a depth interval $\delta z_B/2$ to produce a final set of polynomial coefficients, defining $U_\mathrm{PEDM}$.}

The final current profile may be dependent on the parameters $n_{\mathrm{max}}$, $\delta z_T$, $\delta z_B$, as well as $z_c$, and a method for choosing optimal values of these parameters is necessary. To proceed, we note that when the exact form of the current profile $U(z)$ is considered, the Doppler shifts calculated using (\ref{eq:SJ}) or another suitable approximation method will agree with the measured values save for experimental measurement errors. The process of calculating the Doppler shifts given a prescribed current profile we refer to as the ``forward problem.'' Though the accuracy of (\ref{eq:SJ}) and its finite depth counterpart \citep{skop87} is likely sufficient, we use a direct integration method of arbitrary accuracy due to \citet{li19} to evaluate the Doppler shifts to avoid this unnecessary source of error. We define an RMS difference between the measured Doppler shifts and those calculated by the forward problem ($\tilde{c}_{F,i}$) as
\begin{equation}
\epsilon_{\mathrm{RMS}} = \sqrt{\overline{(\ci-\tilde{c}_{F,i})^2}}\drv{.}{,}
\label{eq:eps}
\end{equation}
\rv{where the overbar represents an average over all wavenumbers.} For accurate evaluation of $\tilde{c}_{F,i}$, the cutoff depth was chosen as $z_c = 2(\mathrm{min}[k_i])^{-1}$ (four times the deepest mapped depth), being set to the water depth in cases where the bottom was shallower than $z_c$. 

Values of $n_{\mathrm{max}}$, $\delta z_T$, and $\delta z_B$ were in practice chosen to minimize $\epsilon_{\mathrm{RMS}}$ to in a sense find the most probable current profile in the presence of experimental noise.


\section{Experimental and Data Analysis Methods}
\label{sec:exp}

\begin{figure}[htb]
\centering
	\includegraphics[width=\textwidth]{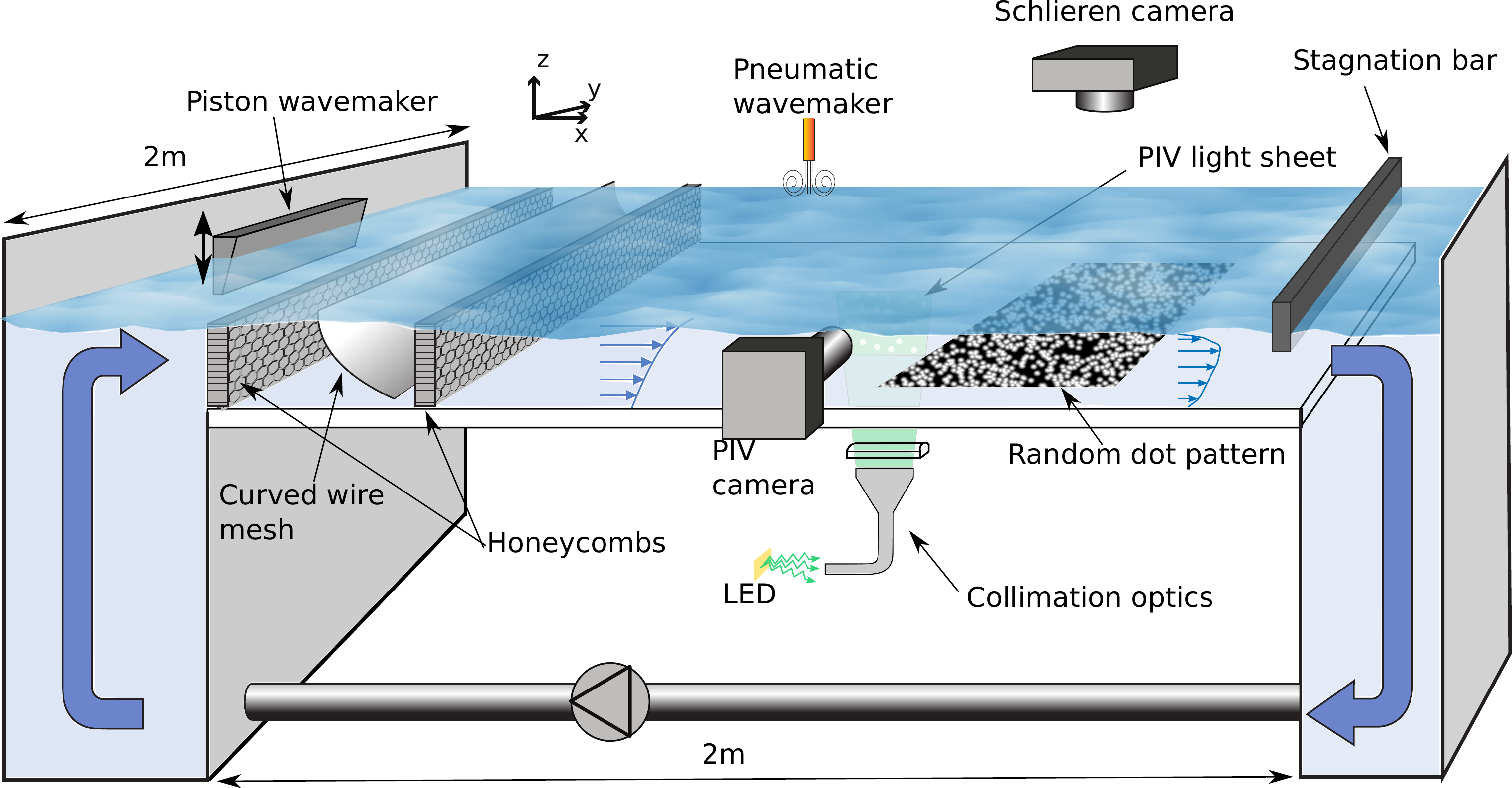}
	\caption{The laboratory setup used for measuring wave spectra in the presence of a controllable background shear flow.}
	\label{fig:lab}
\end{figure}
We test and evaluate the accuracy of the PEDM on experimental data by measuring wave spectra of waves propagating atop a controlled background shear flow generated in a small-scale laboratory setup, shown in Figure~\ref{fig:lab}. The current depth-profile of the shear flow is measured by particle image velocimetry (PIV), which can be used as ``truth" data to compare against the profiles obtained by the PEDM. 

The setup consists of a pump which drives laminar flow over a 2x2 meter transparent plate, where different shear profiles can be obtained by various methods of flow conditioning. One method consists of a sequence of honeycomb structures and a curved wire mesh \citep{dunn07}, which distorts the streamlines of the flow producing a profile with peak velocity at the surface, and decreasing with depth with approximately constant shear. The surface current and near-surface shear strength can be controlled by adjusting the water depth and pump power. It is noted that strong shear near the bottom due to the boundary layer is also created, yet for the depths ($\sim 8 - 10$ cm) and wavelengths we consider the influence of the boundary layer on wave dispersion is negligible. Another method is to make use of a region of flow where the water surface is nearly stagnant (at rest in the laboratory frame of reference) which occurs near the downstream end of the system due to the formation of a Reynolds ridge from surface contaminants \citep{scott82}. The region exhibits strong near-surface shear as the incoming flow dips beneath the stationary viscoelastic surface layer to form a surface boundary layer. The upstream extent of this stagnation region can be increased by the insertion of a horizontal bar in the downstream end as shown in Figure~\ref{fig:lab}. A laboratory coordinate system is defined as shown in Figure~\ref{fig:lab}, with the $x$, $y$ and $z$-axes aligned with the streamwise, spanwise and vertical dimensions respectively.

The depth profile of the shear flow was measured at varying locations in the streamwise and spanwise directions using a planar PIV setup with high power light-emitting diodes (LEDs) as the illumination source similar to the system of \citet{willert10}. Emission from the LED's (Luminus PB-120) was approximately collimated in one dimension to produce a planar light sheet using either a fiber bundle splayed out into a linear array and a cylindrical lens, or a thin rectangular slit mounted above the LED array. The water was seeded with 40 $\mu$m \rv{diameter} polystyrene spheres (Microbeads AS), and particle images were acquired by a camera mounted out of the plane as shown. Image pairs were processed to obtain the streamwise velocity as a function of depth. The setup could be translated to perform measurements at different positions in both horizontal dimensions.

Waves were created using a vertical piston wavemaker mounted at the upstream end of the setup. The wavemaker was run at variable frequencies from 1 to 4 Hz as a function of time, 10 s at each constant frequency in steps of 0.1 Hz, to produce a sufficiently wide spectrum in frequency-wavevector space. The waves were measured using a synthetic Schlieren (SS) method \citep{moisy09}, consisting of a random dot pattern mounted below the transparent bottom plate, and viewed from above by a camera mounted $\sim$ 2 m optical path length from the free surface. The gradient of the free surface, $\nabla \eta(x,y,t) \equiv [\eta_x(x,y,t),\eta_y(x,y,t)]$, can be found by digital image correlation (DIC), comparing camera frames of the dot pattern beneath a perturbed free surface to that of an unperturbed reference frame. Uncertainty in the measured gradients was estimated to be 0.001 based on analysis of images taken with a still water surface. Typical measured root mean squared (RMS) gradients of the waves were between 0.02 -- 0.1 in magnitude, resulting in a relative uncertainty of 5\% or less.

The frequency-wavevector spectrum of the wave gradient field in a 10 s time window roughly corresponding to a given driven wavemaker frequency was calculated as

\begin{equation}
P^l(k_x,k_y,\omega) = |P_x^l(k_x,k_y,\omega)|^2 + |P_y^l(k_x,k_y,\omega)|^2,
\label{eq:spec}
\end{equation}
where $P_x^l$ and $P_y^l$ are the three dimensional discrete Fourier transforms in spatial and temporal dimensions of the surface gradient components obtained directly from the SS method, which are first multiplied with a spatiotemporal windowing filter prior to transformation,

\begin{equation}
 F(x,y,t) = \exp\left[
 -\frac1{2\sigma_m^2}\left(\frac{x^2}{L_x^2}+ \frac{y^2}{L_y^2}+\frac{t^2}{T^2}\right)\right],
\label{eq:filter}
\end{equation}
where $L_x$ and $L_y \sim 0.5$ m are the physical lengths of the spatial domain, $T = 10$ s the extent of the temporal domain, and $\sigma_m = 1/4$. The spatiotemporal domain is assumed to be centered around $\{x,y,t\}=0$ such that $F(x,y,t)$ is peaked in the center of the domain. The spectra $P^l$ for each time window were summed together to produce a single spectrum $P = \sum_l P^l$ containing all frequency spectral components. For the purposes in this work, the fact that the wave spectrum is defined with the free surface gradient instead of the free surface elevation is insignificant, since the gradient field has the same periodicity in space and time as the surface elevation.  

\begin{figure}[htb]
\centering
	\includegraphics{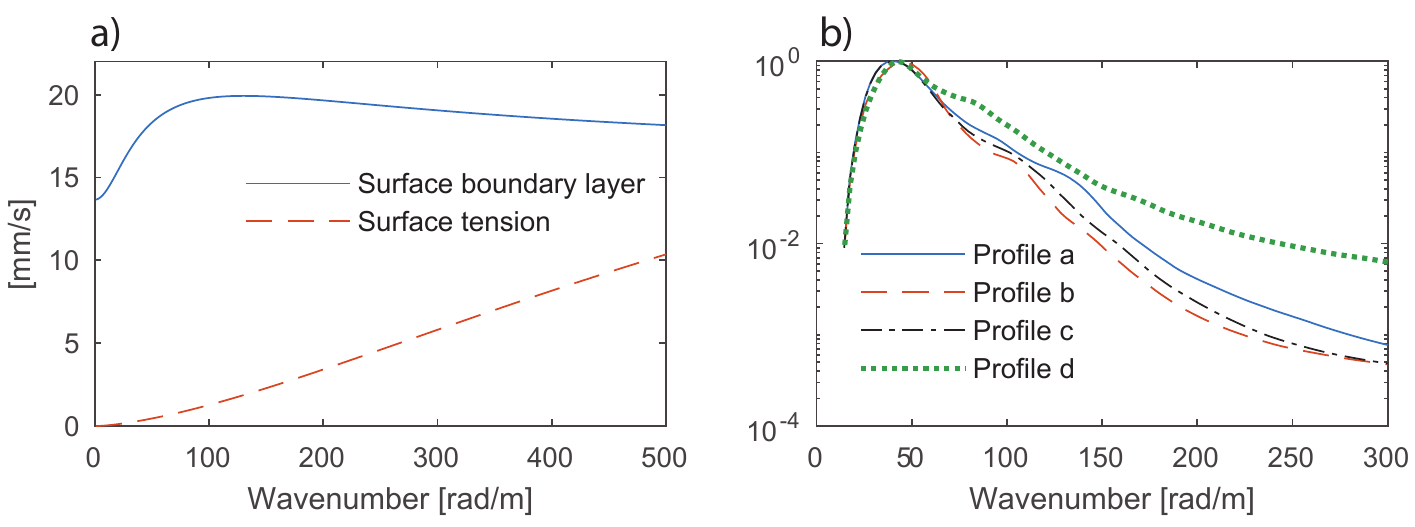}
	\caption{a) Calculated variation of the phase velocity across the streamwise dimension of the measurement area for downstream propagating waves, due to the surface boundary layer development as well as surface tension gradient. b) Azimuthally-averaged gradient spectra $S(k)$ for waves atop the four shear profiles considered here. The spectra are normalized by the peak value.}
	\label{fig:ih_ksp}
\end{figure}

Assuming small wave-steepness, maximum values of the gradient spectrum $P$ are concentrated on the linear dispersion surface $\omega_\mathrm{DR}(k_x,k_y)$, which was assumed to be the sum of two components, a quiescent water term and a term due to the subsurface flow (\ref{eq:dr}). The quiescent water dispersion relation $\omega_0(k)$ is of the form
\begin{equation}
\omega_0(k) = \sqrt{\left(gk + \frac{\sigma}{\rho}k^3\right)\tanh kh},
\label{eq:quies}
\end{equation}
with $g$ the gravitational constant, $\sigma$ the surface tension constant, and $\rho$ the water density. The surface tension coefficient depends on the level of contamination of the water, and was determined by analyzing the wave spectrum recorded with the pump turned off using a pneumatic wavemaker discharging bursts of air at controlled frequencies of 5-10 Hz. A set of frequency-wavenumber pairs $\{k_i,\omega_i\}$ were extracted by finding the peak wavenumber $k_i$ of the spectrum along various directions in $\mathbf{k}$ space for a given frequency $\omega_i$. The set of points was then fit to (\ref{eq:quies}) with $\Gamma \equiv \sigma/\rho$ the fitting parameter. For the stagnation region flows, contaminants become concentrated in the viscoelastic surface layer, and thus a notably  different value of the surface tension coefficient may result when compared to quiescent waters where the contaminants disperse over the whole water channel surface. To obtain a representative value of the surface tension in the stagnation region, we insert horizontal bars dipping just below the surface at the upstream and downstream boundaries of the measurement region and spanning the entire width of the channel in the $y$-direction prior to turning the pump off. The bars prevent the spreading of the surface contaminants over the entire channel region when the pump is turned off. Within the stagnation region there is in fact a gradient in surface tension in the streamwise direction, necessary to balance the surface shear stress of the fluid \citep{harper74}. Using values of the viscosity in clean water and the maximum measured surface shear based on profiles measured by PIV, we estimate the variation of the surface tension coefficient $\sigma$ to be 0.008 $\mathrm{Nm}^{-1}$ across the measurement area, or $8\times 10^{-6}$ $\mathrm{m}^3\mathrm{s}^{-2}$ in the value of $\Gamma$, a relative variation of $\sim 20$\%. 
We assume the measurements of $\Gamma$ using the method described above to be representative of the spatially averaged value within the measurement region. The effect of the inaccuracy thus introduced on our results will be discussed shortly. 

The process here of determining the surface tension coefficient $\Gamma$ is specific to the small-scale laboratory setup, as in most practical cases in the field the length scales of the measured waves are in the regime where surface tension forces can be neglected. In cases when investigating short wavelengths in the ocean \citep[e.g.,][]{laxague17}, a reasonable estimate to the surface tension coefficient and density can be assumed a priori.

Both (\ref{eq:dr}) and (\ref{eq:quies}) describe wave propagation assuming fluid properties ($\Gamma$, $h$, and $\mathbf{U}$) to be invariant across horizontal spatial dimensions. However, for the case of the stagnation region flows, both $\Gamma$ and $U(z)$ vary across the streamwise dimension, due to the surface shear stress balance and the development of the surface boundary shear layer respectively. To quantify the effect these variations have on wave dispersion, we calculate the difference in phase velocities for a wave propagating at the upstream versus downstream ends of the measurement region. For the case of surface tension, we assume $\Gamma$ to vary by $8\times 10^{-6}$ $\mathrm{m}^3\mathrm{s}^{-2}$, and for the surface boundary layer, the difference between the minimum and maximum values of the measured streamwise velocity measured in upstream versus downstream positions for the strongest shear current. The results are shown in Figure~\ref{fig:ih_ksp}a as a function of wavenumber for waves propagating downstream (similar trends occur for upstream propagating waves). The variation of the current profile results in a greater variation in phase velocities ($\sim$ 20 mm/s) across the measurement region compared to surface tension gradients where the variation is $\leq$10 mm/s for the wavenumber range shown. The values in Figure~\ref{fig:ih_ksp}a place a bounds on potential variations and uncertainties of the extracted wave Doppler shifts $\ctil(k)$, though it is expected that Doppler shifts will be representative of the spatially averaged values across the measurement region. For current profiles produced with the curved mesh configuration, significantly less variation across the measurement region is expected once the shear profile has reached a stable state within the measurement area, and there is in this case no streamwise gradient in surface tension.

The Doppler shift velocities as a function of wavenumber were extracted by analyzing the gradient spectrum spectrum $P$. The range of wavenumbers to consider was chosen based on the azimuthally-averaged wave number spectrum:

\begin{equation}
S(k) = \int_0^{2\pi}\int_0^\infty\mathrm{d}\omega\mathrm{d}\theta\left(|P_x(\bk,\omega)|^2 + |P_y(\bk,\omega)|^2\right),
\label{eq:kspec}
\end{equation}
where $\bk = \{k\cos\theta,k\sin\theta\}$ is defined in polar coordinates here where $\theta$ is the angle in the $x,y$ plane from the positive $x$-axis. The spectra of waves atop the four current profiles considered here are shown in Figure~\ref{fig:ih_ksp}b as a function of wavenumber, scaled by their maximum value. The wavenumber range for extraction of Doppler shifts was chosen to be wavenumbers where $S(k)$ was greater than 0.1 of the peak value for wavenumbers less than the peak value, and greater than 0.02 of the peak value for wavenumbers larger than the peak value. The minimum wavenumber was $\sim$20 $\rev{\mathrm{rad}}\cdot\mathrm{m}^{-1}$ for all profiles, and the maximum between $\sim$120-190 $\rev{\mathrm{rad}}\cdot\mathrm{m}^{-1}$. A set of wavenumbers $k_i$ was specified spanning minimum to maximum values in steps of $2\pi/(10L_x)$.

\begin{figure}[htb]
\centering
	\includegraphics{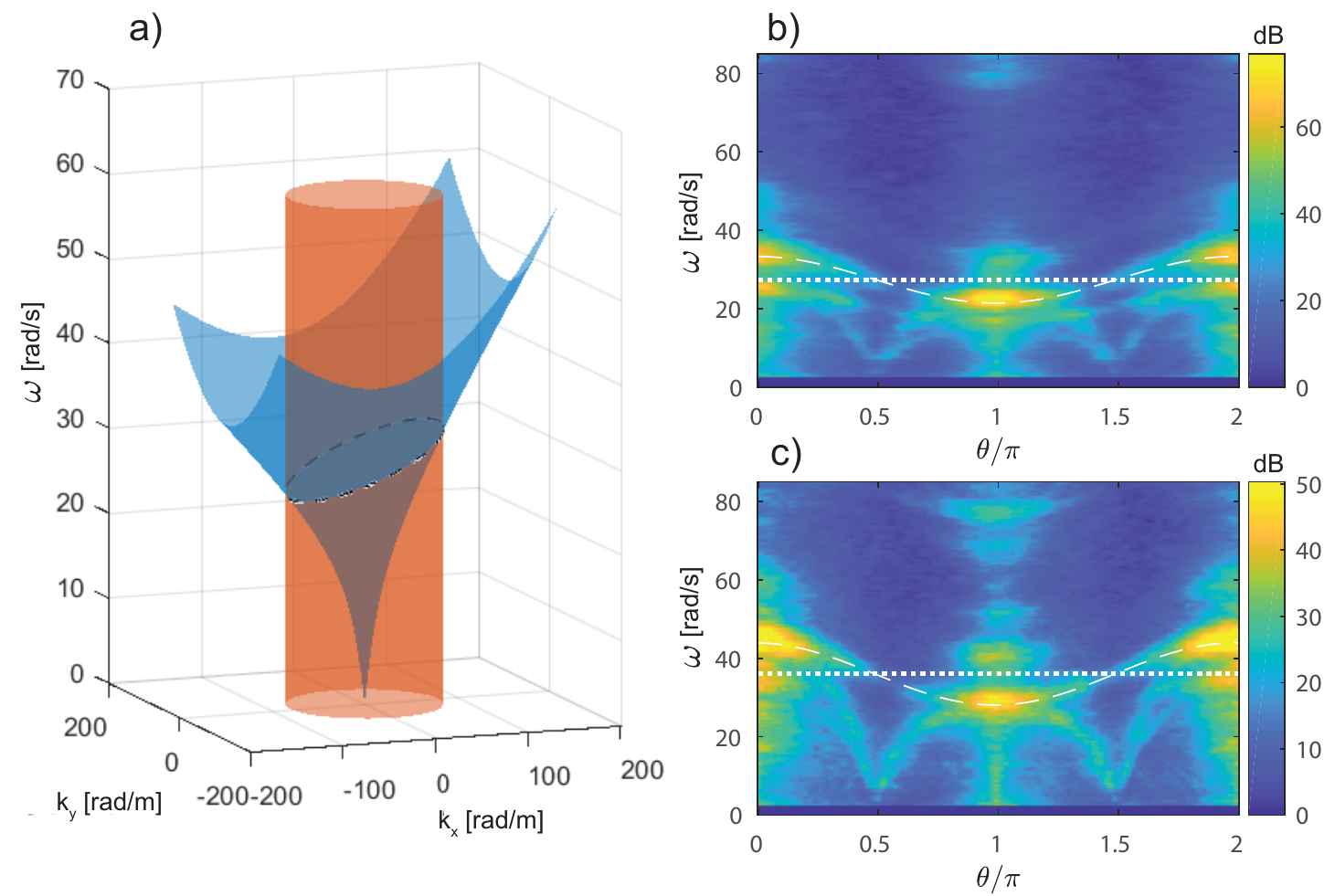}
	\caption{a) An illustration of the dispersion surface (\ref{eq:dr}) and a cylindrical surface of constant wavenumber, with the intersection shown as the dashed curve. b)-c) Values of the \rev{modified} gradient spectrum $P\rev{_i^\prime}(k_i\cos\theta,k_i\sin\theta,\omega)$ on the surface of constant wavenumber for $k_i$ = \drev{59}{75} and \drev{155}{125} $\rev{\mathrm{rad}}\cdot\mathrm{m}^{-1}$, respectively. The frequency as a function of $\theta$ reflecting the extracted Doppler shifts is shown as the dashed curve, while the frequency in quiescent waters is shown as the dotted line.}
	\label{fig:nsp}
\end{figure}

For each wavenumber $k_i$, Doppler shift velocities were extracted by considering the signal $P(\bk,\omega)$ on a cylindrical surface of constant wavenumber magnitude $k_i$ in $(\bk,\omega)$ space, and using an NSP method \rev{\citep{huang16,huang12,serafino10}}. The cylindrical surface as well as the dispersion surface from (\ref{eq:dr}) is shown in Figure~\ref{fig:nsp}a for the case of a depth-uniform current. The method works to effectively determine the frequency of intersection $\omega_{\mathrm{DR}}(k_i,\theta)$ as a function of $\theta$ between the cylindrical surface and the dispersion relation surface (which corresponds to peak values of $P$)\rev{, where the wavevector arguments of $\omega_{\mathrm{DR}}$ are expressed in polar coordinates}. From (\ref{eq:dr}), it is apparent that in quiescent waters ($\ctil(k) = 0$) the frequency of intersection is independent of azimuth angle, whereas in the presence of a current there is an additional oscillating component with amplitude and phase determined by $\ctil(k)$, as seen in Figure~\ref{fig:nsp}a as the dashed curve. \dl{Example values of $P$ on cylindrical surfaces of constant wavenumber are shown in Figure~\ref{fig:nsp}b and c for $k_i$ = 59 and 155 $\mathrm{rad}\cdot\mathrm{m}^{-1}$ respectively, as a function of $\theta$ and $\omega$ for waves atop a shear current.  In both cases, the peak frequency as a function of $\theta$ displays a clear oscillatory trend due to the presence of shear as expected.}

We proceed by finding Doppler velocity components $\tilde{c}_{x,i}$ and $\tilde{c}_{y,i}$ at wavenumber $k_i$. First, we define a characteristic function

\begin{equation}
G_i(\omega,\theta,\tilde{c}_{x,i},\tilde{c}_{y,i}) = \exp\left[\frac{(\omega-\omega_{\mathrm{DR}}(\rev{k_i,\theta}))^2}{4a(\theta)}\right],
\end{equation}
where $a = (\sigma_mT)^{-2}$ is defined based on the Gaussian width in Fourier space given the applied spatial Gaussian filter $F$ defined in equation (\ref{eq:filter}). Dependence on $\tilde{c}_{x,i}$ and $\tilde{c}_{y,i}$ is implicitly included in $\omega_{\mathrm{DR}}$.  In addition, \dl{to improve the sensitivity to currents }we consider the second harmonic spectral components $\{2\bk,2\omega\}$ and define a modified spectrum
\begin{equation}
P_i^\prime(\theta,\omega) = \rev{10}\log P(k_i\cos\theta,k_i\sin\theta,\omega) + \rev{10}\log P(2k_i\cos\theta,2k_i\sin\theta,2\omega),
\label{eq:p2h}
\end{equation}
where $P_i^\prime$ is \rev{then} scaled such that the minimum value is zero. The signal at the higher harmonic is due to the weak non-linearity of the measured surface waves as well as non-linearities in the SS measurement system \citep{senet01}. \rev{Assuming the spectral peak associated with the second harmonic has comparable spectral width as the fundamental harmonic, the contribution to the peak from the second term in (\ref{eq:p2h}) would have a smaller width in $\theta$-$\omega$ space due to the factor two in the argument of $P$. Including the second harmonic may thus increase the sensitivity to currents by making the peak of $P_i^\prime$ more localized.} \rev{Example values of $P_i^\prime$ on cylindrical surfaces of constant wavenumber are shown in Figure~\ref{fig:nsp}b and c for $k_i$ = 75 and 125 $\rev{\mathrm{rad}}\cdot\mathrm{m}^{-1}$ respectively, as a function of $\theta$ and $\omega$ for waves atop a shear current.  In both cases, the peak frequency as a function of $\theta$ displays a clear oscillatory trend due to the presence of shear as expected.}

We find the Doppler shift velocities by maximizing the scalar product $N$ between $G$ and $P_i^\prime$:

\begin{equation}
N(\tilde{c}_{x,i},\tilde{c}_{y,i}) = \frac{\langle G(\omega,\theta,\tilde{c}_{x,i},\tilde{c}_{y,i})P_i^\prime(\theta,\omega)\rangle}{\langle G\rangle\langle P^\prime\rangle},
\end{equation}
where $\langle ...\rangle$ indicates a double integral over $\theta$ and $\omega$ (the same integral as (\ref{eq:kspec})). To avoid 
local maxima other than those associated with the dispersion relation, $N$ is first evaluated on a grid of points spanning expected values of the Doppler shift velocity components, with the Doppler shifts corresponding the maximum value of $N$ used as an initial guess for further optimization. The resulting curves $\omega_{\mathrm{DR}}(k_i\cos\theta,k_i\sin\theta)$ from the fitting routine are shown as the dashed lines in Figure~\ref{fig:nsp}b-c. The dotted lines show the frequency in quiescent waters. As can be seen, there is a distinct departure in the peak signal as a function of angle that is captured by the NSP fit, but inconsistent with the quiescent water frequency as it should be. The Doppler shifts as a function of wavenumber are expected to display a smooth functionality based on (\ref{eq:SJ}), and values were removed using an outlier filter. Both components were fit to a first order polynomial to produce functions $\tilde{c}_x^O(k)$ and \drev{$\tilde{c}_x^O(k)$}{$\tilde{c}_y^O(k)$}. \drev{Doppler shifts where $(\tilde{c}_{x,i}-\tilde{c}_x^O(k_i))/\sqrt{\tilde{c}_x^O(k_i)^2 + \tilde{c}_y^O(k_i)^2}>0.25$}{Outliers were identified by considering the set $\{\tilde{c}_{x,i}-\tilde{c}_x^O(k_i)\}$} (and the equivalent for the $y$-direction) \drev{were removed from the data set}{and data lying more than 1.5 times the interquartile range below the first quartile and the same interval above the third quartile were removed}.

\section{Results and Discussion}
\label{sec:expres}

\begin{figure}[htb]
\centering
	\includegraphics{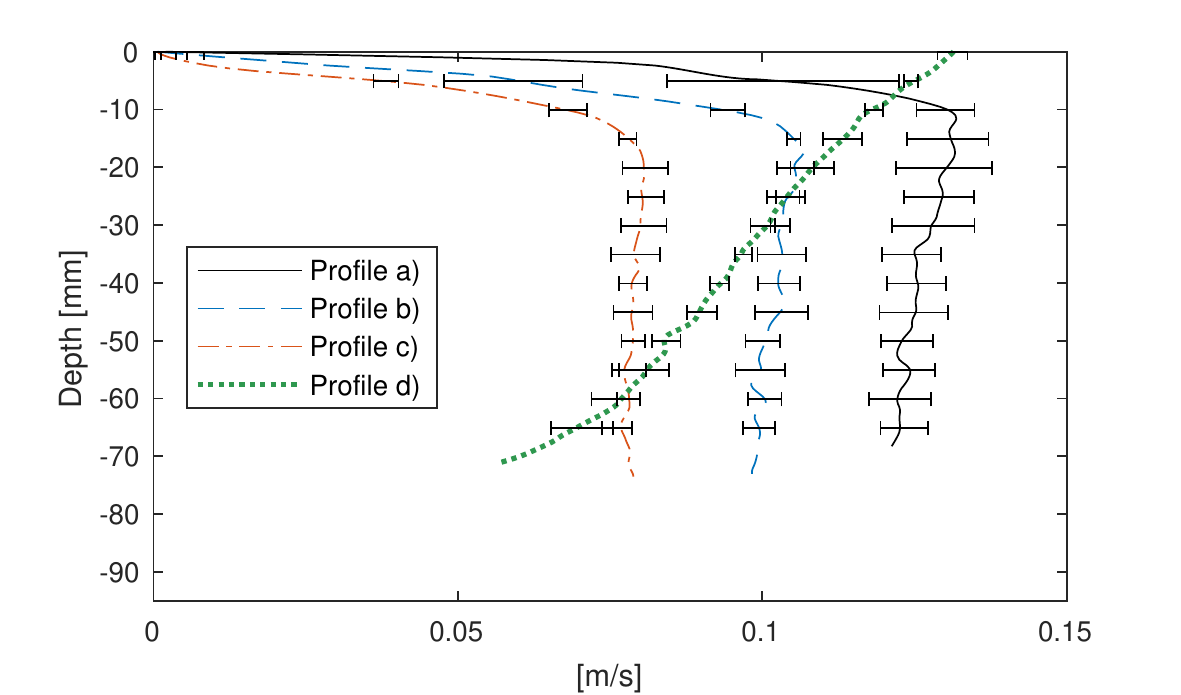}
	\caption{Current profiles measured by PIV. The error bars denote the range of measured velocities at different streamwise and spanwise positions within the wave measurement area. \rev{The water depth was 95 mm for profiles a)-c) and 80 mm for Profile d).} }
	\label{fig:upiv}
\end{figure}

To validate and examine the accuracy of the PEDM, we apply it to Doppler shifts measured in the laboratory setup with the current profile measured by PIV used as ``truth" measurements to compare against. We consider experimental data for waves atop 4 different shear flows, referred to as profiles a-d), shown in Figure~\ref{fig:upiv}. Profiles a-c) are in the stagnation region at different flow rates which lead to varying near surface shear strengths and curvature. Profile d) was produced using the curved wire mesh, and had weaker surface shear strength and near-constant vorticity with depth. The parameters including the measured surface tension coefficient $\Gamma$ are given in Table 1. The velocity was not measured for the bottom $\sim$1-2 cm depth where the bottom boundary layer was located.

 \begin{table}
 \label{t:prof}
 \caption{Summary of properties for the four laboratory current profiles. }
 \centering
 \begin{tabular}{l c c c c}
 \hline
  Profile  & Flow Type & Water Depth & Flow rate & $\Gamma$  \\
  & & [mm] & $[\mathrm{m}^3/$s$]$ & $\times 10^{-5}$ $[\mathrm{m}^3\mathrm{s}^{-2}]$ \\
 \hline
     a  & Stagnation region & 95 & 0.021 & $3.8\pm 0.05$   \\
     b  & Stagnation region & 95 & 0.017 & $2.8 \pm 0.1$    \\
     c  & Stagnation region & 95 & 0.014 & $2.7 \pm 0.1$  \\
     d  & Curved mesh & 80 & 0.014 & $6.7\pm 0.1$   \\
 \hline
 \multicolumn{2}{l}{}
 \end{tabular}
 \end{table}

\begin{figure}[htb]
\centering
	\includegraphics{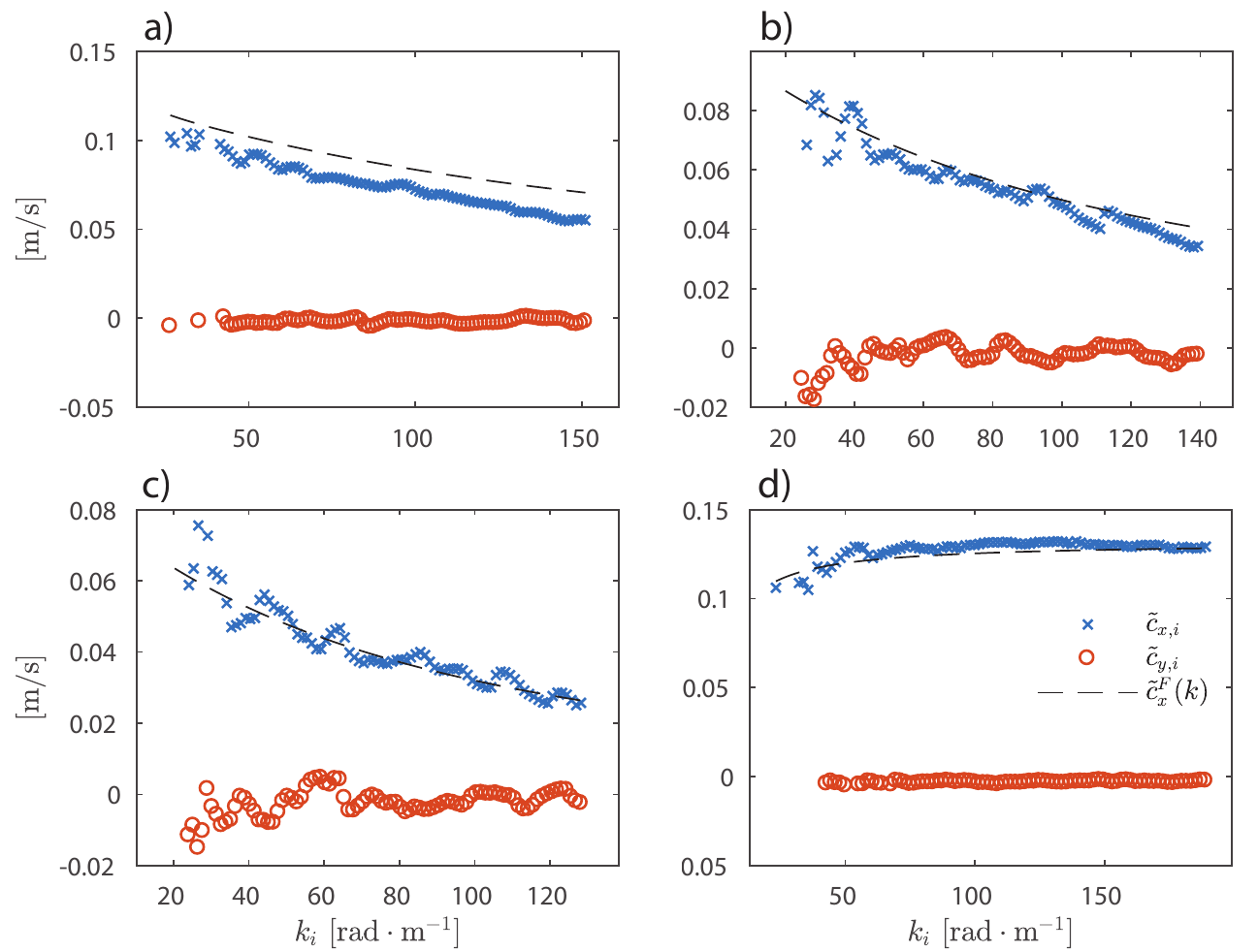}
	\caption{Experimentally measured Doppler shifts as a function of wavenumber, for current profiles a-d) shown in panels a-d) respectively. The x-marks are the $x$-component of the Doppler shifts, while the circles are $y$-component. Calculated Doppler shifts from theory using the current profile as measured by PIV are shown as the dashed curves. }
	\label{fig:ctil_exp}
\end{figure}

The measured Doppler shifts using the NSP method described in section \ref{sec:exp} for the four profiles are shown in Figure~\ref{fig:ctil_exp}. Doppler shifts $\tilde{c}_x^F(k)$ calculated with theory assuming the measured PIV profile are shown as the dashed curves. As no mean flow in the $y$-direction was expected, the \rv{true values of the} $y$-components of the Doppler shifts $\tilde{c}_{y,i}$ were assumed to be zero \drv{in theory}{at all depths}. Differences between experiment and theory are $\leq 1$ cm/s over most wavenumbers, except for a 1-2 cm/s bias for profile a). The reason for the bias is not known, yet could be a result of a greater streamwise variation in the shear profile given that the pump power was greatest for this profile. 

\begin{figure}[htb]
\centering
	\includegraphics{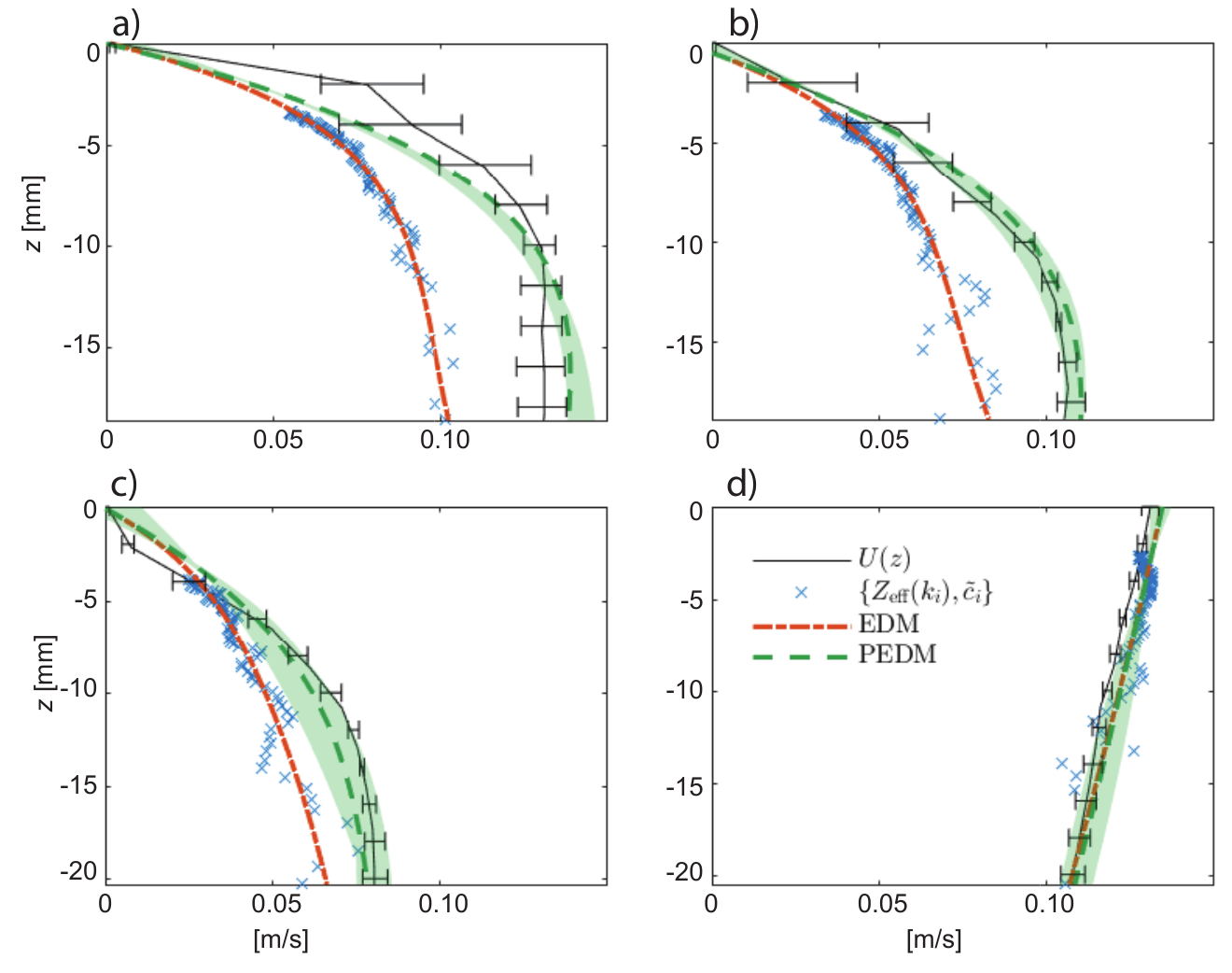}
	\caption{Results of the PEDM applied to the $x$-components of the measured Doppler shifts for profiles a\del{)}-d). The profile measured by PIV $U(z)$ is shown as the solid curve, with error bars denoting the range of measured velocities at different streamwise and spanwise positions within the wave measurement area. The initial mapped profile $U_{\mathrm{EDM}}(z)$ is also shown for comparison. The vertical depth-axis extends down to the greatest mapped depth, and the legend applies to all panels. The shaded regions are bounds on the current strength based on all PEDM profiles using parameter combinations ($n_{\mathrm{max}}$, $\delta z_T$, and $\delta z_B$) where $\epsilon\rv{_\mathrm{RMS}}$ was within 10\% of the minimum value.
	}
	\label{fig:pedm_u}
\end{figure}

The PEDM was implemented as described in section \rev{\ref{sec:pedm_impl}} for each component of the Doppler shifts separately. Current profiles $U_{\mathrm{PEDM}}(z)$ were calculated with 900 combinations of the parameters $n_{\mathrm{max}}$, $\delta z_T$, and $\delta z_B$: $10$ values of $\delta z_T$ and $\delta z_B$ each, ranging from $0.5$-$4$ mm and $1$-$20$ mm, respectively, and $9$ values of $n_{\mathrm{max}}$ ranging from $2$-$10$. For each combination, the RMS difference $\epsilon_{\mathrm{RMS}}$ between the measured Doppler shifts and those from the forward problem with $U_{\mathrm{PEDM}}(z)$ was evaluated. Profiles where the initial polynomial fit $U_{\mathrm{EDM}}(z)$ was not monotonic were discarded. The combination of parameters that gave the lowest value of $\epsilon_{\mathrm{RMS}}$ were used to produce a profile that was presumed to be the most probable estimate. 

The monotonic assumption was based on the fact that the Doppler shifts (of which $U_{\mathrm{EDM}}(z)$ is based) can be viewed as a weighted average of the current depth-profile, thus resulting in a large degree of smoothing of oscillations in the true profile when considering $U_{\mathrm{EDM}}(z)$ obtained from the mapped depths. Over a finite range of wavenumbers, it is assumed that the true Doppler shifts are monotonic for most all realistic current profiles, and that profiles $U_{\mathrm{EDM}}(z)$ that are not monotonic result from errors in the Doppler shifts. It is however important to note that the monotonic assumption here does not also constrain the profile $U_{\mathrm{PEDM}}(z)$, given the scaling of the polynomial coefficients.

\rev{The process of calculating $U_{\mathrm{PEDM}}(z)$ profiles and evaluating $\epsilon_{\mathrm{RMS}}$ for the 900 combinations of PEDM parameters with roughly 100 wavenumber-Doppler shift pairs took approximately 6 minutes on an Intel\textregistered Core\textsuperscript{TM} i7-4770 3.40 GHz processor with 32 GB of RAM. However, the vast majority of time was spent evaluating $\epsilon_{\mathrm{RMS}}$ using the direct integration method. It is noted that for cases where all wavenumbers can be assumed to be in deep water and the approximation accuracy of (\ref{eq:SJ}) is deemed sufficient, (\ref{eq:pedm_ctil}) may be used to evaluate $\tilde{c}_{F,i}$ directly from the PEDM polynomial coefficients. When using (\ref{eq:pedm_ctil}), the same process took only 16 s.} 

The results of applying the PEDM to the $x$-components of the Doppler shifts for the four profiles are shown in Figure~\ref{fig:pedm_u}. The black curve denotes the current profile as measured by PIV, the average over the spatial locations within the wave measurement area with the error\del{s} bars denoting the maximum and minimum values measured by PIV over the spatial locations. Profiles $U_{\mathrm{EDM}}(z)$ and $U_{\mathrm{PEDM}}(z)$ using the optimal set of parameters are shown as the dash-dotted and dashed curves respectively, along with the mapped Doppler shifts. For profiles a-c), the PEDM is a clear improvement over the EDM with notably increased accuracy over most all depths. Given the relatively strong curvature of the profiles, the assumption of a linear profile that was inherent in the mapping function is not valid here, and the mapped Doppler shifts deviate notably compared to the measured current profile. The deviation is greatest for profile a) and successively decreases for profiles b) and c) which is expected based on the weakened curvature of these profiles. For profile d) where the true profile has near-constant vorticity, the assumption of a linear profile is largely valid and the PEDM offers negligible improvement in accuracy over the EDM as may be expected. The shaded regions are discussed shortly.

To evaluate the improvement in accuracy of the PEDM, we calculate the depth-integrated RMS difference $\Delta U_{\mathrm{RMS}}$ between $U_{\mathrm{PEDM}}(z)$ or $U_{\mathrm{EDM}}(z)$ and the profile measured by PIV over the range of mapped depths. The results are summarized in Table 2, along with the optimal PEDM parameters for each profile. The ratio shown in the rightmost column is the degree of improvement in accuracy achieved by the PEDM relative to the EDM. An improvement of $>3\times$ is achieved for profiles a-c), with a maximum improvement of \drev{13.7}{5.1}$\times$ for profile b). \dl{We consider this value to be potentially fortuitous, yet profiles a) and c) confirm the significant improvement in accuracy that may be expected for profiles with high degrees of near-surface curvature. }\rv{For profile d), the PEDM is marginally less accurate than the EDM, yet the absolute value of $\Delta U_{\mathrm{RMS}}$ remains small compared to the other profiles.} For all profiles, the PEDM achieves a depth-integrated RMS absolute accuracy $<10$ mm/s relative to the PIV profiles.

 \begin{table}
 \label{t:t1}
 \caption{Summary of the optimal parameters and results of the PEDM applied to $x$-components of experimentally measured Doppler shifts. }
 \centering
 \begin{tabular}{l c c c c c c}
 \hline
  Profile  & $n_{\mathrm{max}}$ & $\delta z_T$ & $\delta z_B$ & $\Delta U_{\mathrm{RMS}}^{\mathrm{EDM}}$ & $\Delta U_{\mathrm{RMS}}^{\mathrm{PEDM}}$  & $\frac{\Delta U_{\mathrm{RMS}}^{\mathrm{EDM}}}{\Delta U_{\mathrm{RMS}}^{\mathrm{PEDM}}}$  \\
  & & [mm] & [mm]&[mm/s] & [mm/s] & \\
 \hline
     a  & 8 & 0.5 & 17.9 & \drev{32.7}{34.2} & \drev{9.3}{8.9} & \drev{3.5}{3.8}   \\
     b  & 8 & \drev{1.3}{0.5} & 3.1 & \drev{22.1}{21.9} & \drev{1.6}{4.3} & \drev{13.7}{5.1}  \\
     c  & 10 & 1.7 & 5.2 & 14.2 & \drev{3.8}{3.0} & \drev{3.7}{4.8}   \\
     d  & \drev{2}{3} & \drev{0.5}{4.0} & \drev{13.7}{7.3} & \drev{4.1}{3.0} & \drev{3.5}{3.4} & \drev{1.2}{0.9}   \\
 \hline
 \multicolumn{2}{l}{}
 \end{tabular}
 \end{table}

\begin{figure}[htb]
\centering
	\includegraphics{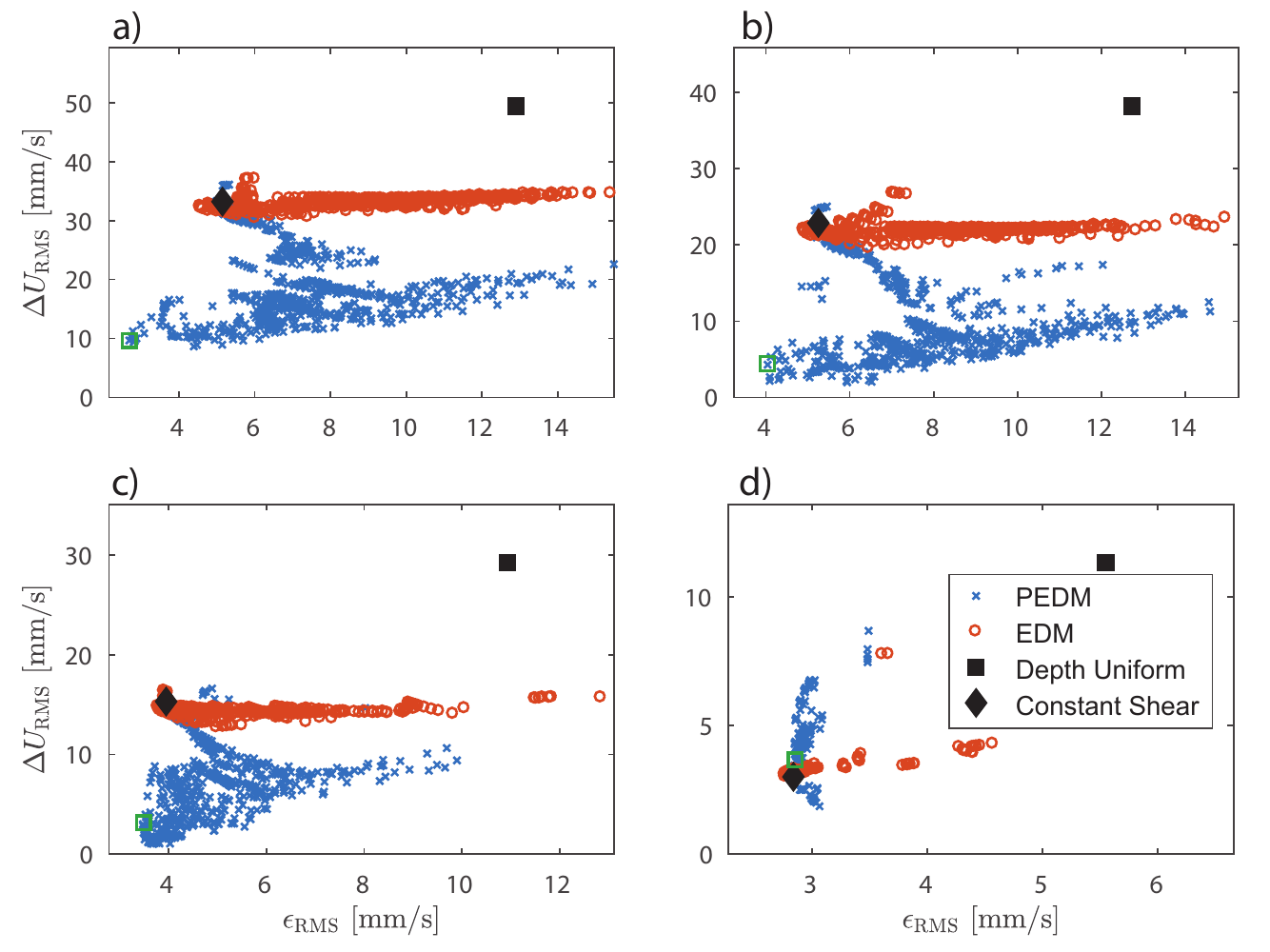}
	\caption{$\Delta U_{\mathrm{RMS}}$ and $\epsilon_\mathrm{RMS}$ for all PEDM parameter combinations for current profiles a-d). Resulting profiles assuming depth-uniform flow ($n_\mathrm{max} = 0$) and constant shear ($n_\mathrm{max} = 1$) are also shown. The legend applies to all panels. The open squares (green) mark the parameter combination with minimum $\epsilon_\mathrm{RMS}$ that was used for the $U_\mathrm{PEDM}(z)$ curves shown in Figure~\ref{fig:pedm_u}.}
	\label{fig:pedm_eps}
\end{figure}

\subsection{Dependence on PEDM parameters}

By using the combination of parameters $n_{\mathrm{max}}$, $\delta z_T$, and $\delta z_B$ that give the minimum $\epsilon_\mathrm{RMS}$ value, the values are thus set algorithmically during the running of the PEDM ``algorithm" rather than as a required input determined prior to it. Thus from a user perspective the method is made effectively parameter free as we will now explain. It is noted that the same parameters are necessary in the use of the EDM as well, in creating a smooth velocity profile to fit the set of mapped Doppler shifts. 

We examine the dependence of the results on the choice of the PEDM parameters by calculating $\Delta U_{\mathrm{RMS}}$ for each combination of parameters for both $U_\mathrm{EDM}(z)$ and $U_\mathrm{PEDM}(z)$, and plotting $\Delta U_{\mathrm{RMS}}$ against $\epsilon_{\mathrm{RMS}}$ as is shown in Figure~\ref{fig:pedm_eps} for the four current profiles. Also shown are results assuming a depth-uniform profile ($n_\mathrm{max} = 0$) and constant shear ($n_\mathrm{max} = 1$) which are independent of the choice of $\delta z_T$ and $\delta z_B$. It is noteworthy that $\Delta U_{\mathrm{RMS}}$ cannot be evaluated in realistic situations where ``truth'' measurements do not exist, so a criteria for choosing the optimal set of PEDM parameters to achieve a small value of $\Delta U_{\mathrm{RMS}}$ is desired based on metrics such as $\epsilon_{\mathrm{RMS}}$ that may be readily evaluated purely from the wave spectral data. The parameter combinations resulting in the minimum value of $\epsilon_\mathrm{RMS}$ are outlined with the open green squares in Figure~\ref{fig:pedm_eps}, corresponding to the profiles $U_\mathrm{PEDM}$ shown in Figure~\ref{fig:pedm_u}.

Ideally, there would be a strong correlation between small values of $\epsilon_{\mathrm{RMS}}$, which can be calculated from the experimental data only, and $\Delta U_{\mathrm{RMS}}$ for which it is our goal to minimize.  In Figure~\ref{fig:pedm_eps}a-b) for the profiles with the strongest curvature, there is noticeable correlation for the smallest values of $\epsilon_{\mathrm{RMS}}$. For those cases, various values of $\epsilon_{\mathrm{RMS}}$ all yield values of $\Delta U_{\mathrm{RMS}}$ that are a significant improvement over the EDM cases (shown as the circles). It is notable that for profile b) where the PEDM profile with lowest value of $\epsilon_\mathrm{RMS}$ yielded a \drev{13.7}{5.1}$\times$ reduction in $\Delta U_{\mathrm{RMS}}$ relative \rev{to} the EDM, other points near the minimum $\epsilon_\mathrm{RMS}$ value still give a $\sim$3X or greater improvement in accuracy (the same being true for profile a)). For profiles c) and d), there is significantly less correlation between $\epsilon_\mathrm{RMS}$ and $\Delta U_{\mathrm{RMS}}$. Nonetheless, for profile c), the minimum value of $\epsilon_\mathrm{RMS}$ yields a value of $\Delta U_{\mathrm{RMS}}$ that is notably less than that of the EDM and constant shear case. For profile d) there is no significant difference in $\Delta U_{\mathrm{RMS}}$ between the EDM, PEDM, and constant shear cases considering the smallest values of $\epsilon_\mathrm{RMS}$, which may be expected given the approximately linear form of the current profile. For all cases, there is a distinct improvement in accuracy relative to the depth-uniform assumption. In addition, for all profiles the EDM displayed a similar level of accuracy relative to the case of constant shear, which is reasonable given that the same assumption was inherent to the EDM. Furthermore, profiles a) and b) with the greatest degree of curvature display the largest improvement over the constant shear case considering the lowest values of $\epsilon_\mathrm{RMS}$.

Choosing the optimal set of PEDM parameters based on $\epsilon_\mathrm{RMS}$ in a sense can be considered to yield the most probable current profile, i.e. the profile that agrees to the greatest degree with the experimentally measured Doppler shifts. However, given experimental noise it is useful to examine the variation in current profiles for parameter combinations that yield values of $\epsilon_\mathrm{RMS}$ near the minimum value, as those profiles may be considered nearly as probable. We calculate the bounds on the range of current values as a function of depth considering all profiles where $\epsilon_\mathrm{RMS}$ is within 10\% of the minimum value, and show these bounds as the shaded regions in Figure~\ref{fig:pedm_u}. For the stagnation region profiles a-c), the spread is narrowest for a-b) which may be expected based on the stronger correlation between $\epsilon_\mathrm{RMS}$ and $\Delta U_\mathrm{RMS}$ as shown in Figure~\ref{fig:pedm_eps}\rv{,} where small values of $\epsilon_\mathrm{RMS}$ yield a smaller spread in the values of $\Delta U_{\mathrm{RMS}}$. For profile c), the spread in $\Delta U_{\mathrm{RMS}}$ is much greater, and less accurate profiles with near constant shear are included. As Figure~\ref{fig:pedm_eps}c shows\rv{,} the lowest value of $\epsilon_\mathrm{RMS}$ is much closer to that of the EDM even though the improvement in $\Delta U_\mathrm{RMS}$ is very significant. Had the threshold for the shaded region been set lower, the least good, near-linear profiles would be excluded.

Another potential reason for the increased spread in profile c) is the fact that the measured Doppler shifts appear \rv{slightly} less smooth as a function of wavenumber when compared to profiles a-b). Furthermore, for profile a) where the measured Doppler shifts displayed a bias relative to those calculated from theory yet are relatively smooth as a function of wavenumber, the PEDM results in a very narrow spread around the most probable current profile that also has a corresponding bias towards reduced current strength near the surface compared to the PIV profile.

\begin{figure}[htb]
\centering
	\includegraphics{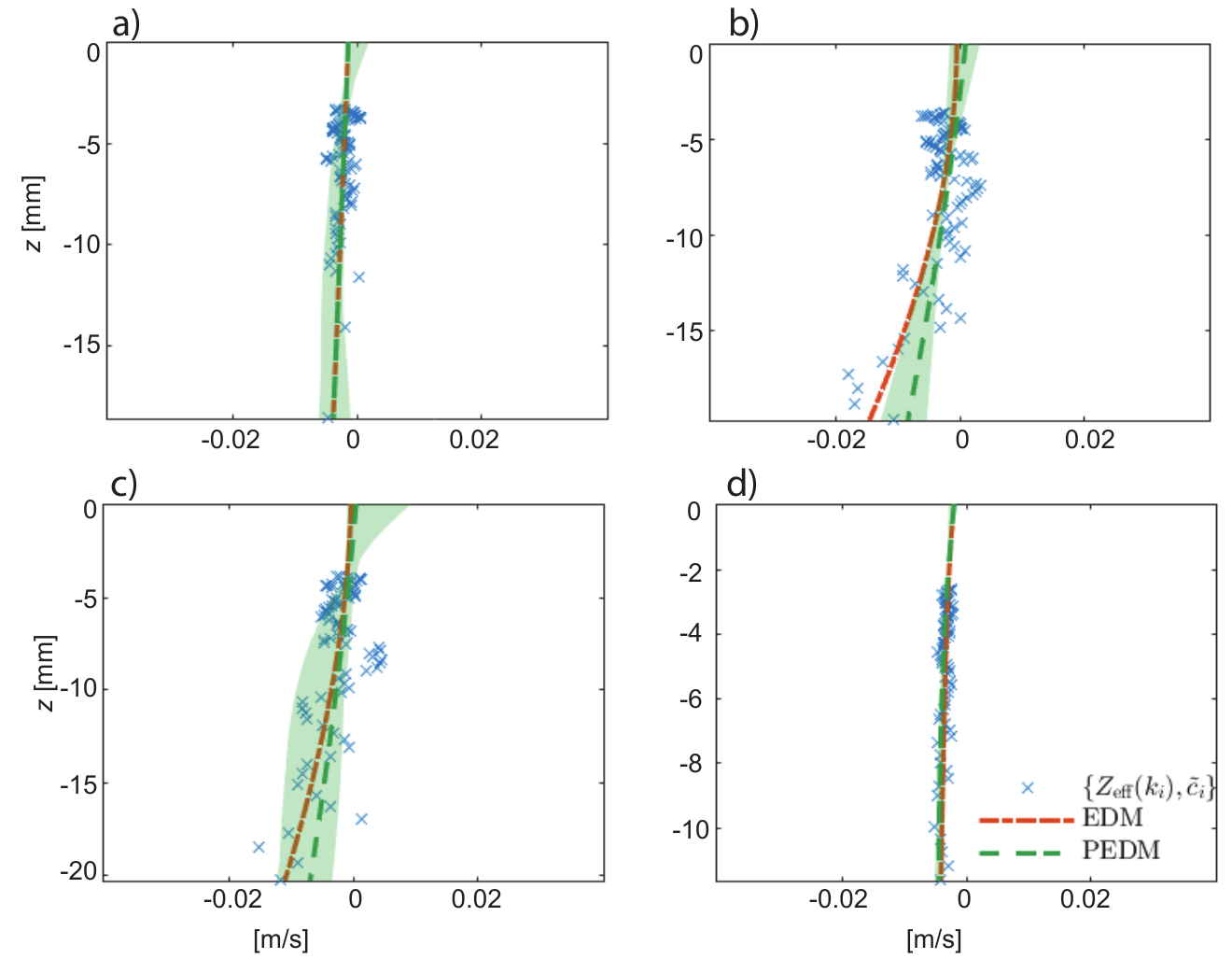}
	\caption{Same as Figure~\ref{fig:pedm_u}, for the $y$-components of the Doppler shifts. For this spanwise direction, the current was assumed to be zero for all depths (not measured).}
	\label{fig:pedm_uy}
\end{figure}

\begin{figure}[htb]
\centering
	\includegraphics{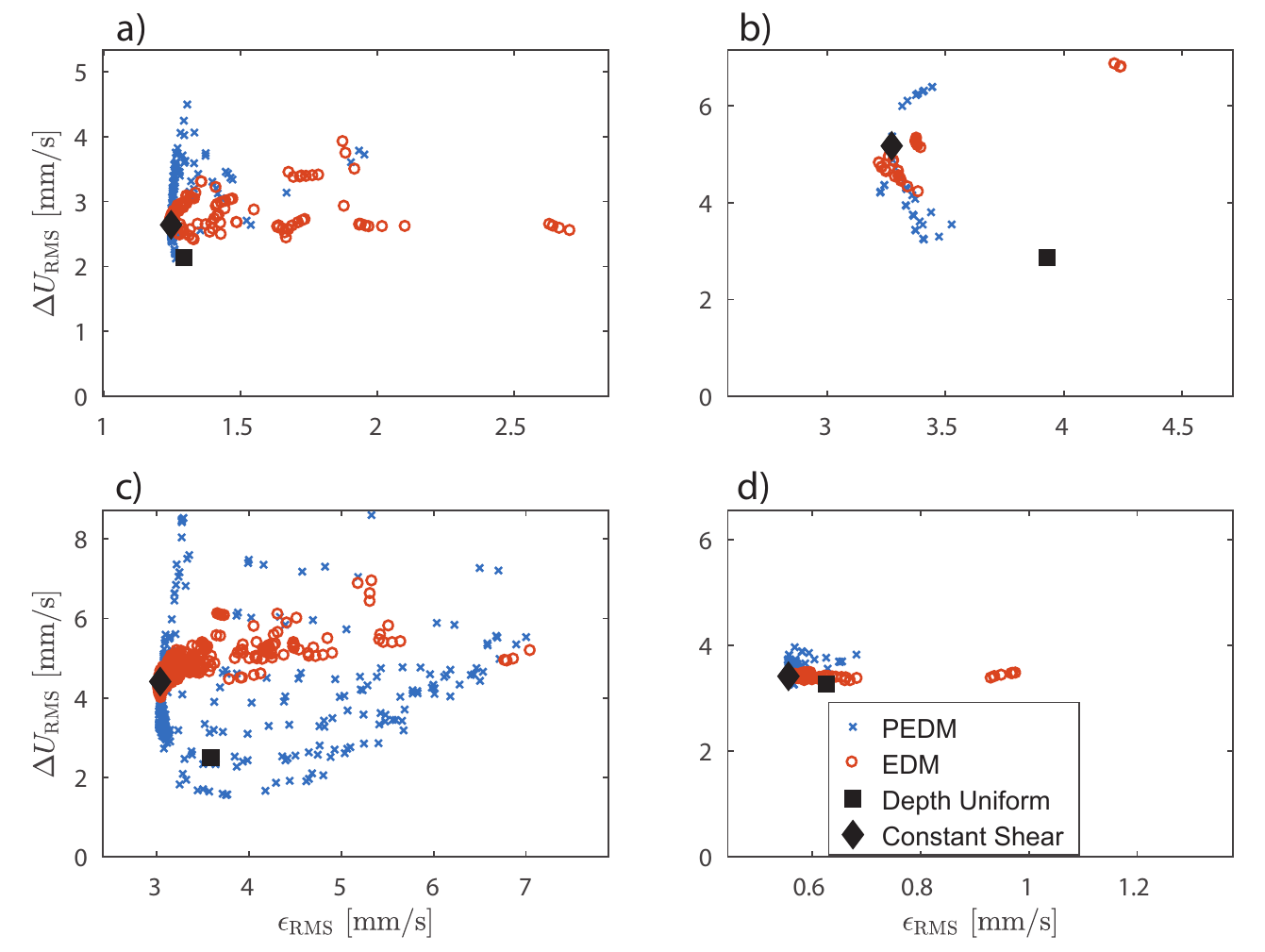}
	\caption{Same as Figure~\ref{fig:pedm_eps}, for the $y$-components of the Doppler shifts.}
	\label{fig:pedm_epsy}
\end{figure}

The same procedure and data analysis is applied to the $y$-components of the measured Doppler shifts and shown in Figures \ref{fig:pedm_uy} and \ref{fig:pedm_epsy}. As there was expected to be no current in this direction for all cases, the results represent the case of a depth-uniform profile in a moving reference frame. As expected, there is negligible improvement in accuracy using the PEDM relative to the EDM. The results serve as further important confirmation that the PEDM results do not deviate significantly from the results of the EDM in cases where the assumptions of a linear profile are valid. As shown in Figure~\ref{fig:pedm_epsy}, assumption of constant shear results in roughly the minimum value of $\epsilon_\mathrm{RMS}$, with only a slight increase in $\Delta U_{\mathrm{RMS}}$ relative to the depth-uniform current assumption. Due to experimental noise, results for both the EDM and PEDM result in slightly sheared current profiles, yet absolute values of $\Delta U_{\mathrm{RMS}}$ remain $<$1 cm/s for all parameter combinations in the vicinity of the minimum $\epsilon_\mathrm{RMS}$ value. Note that the range of values of the horizontal current strength axis in Figure~\ref{fig:pedm_uy} is reduced compared to Figure~\ref{fig:pedm_u}.

%

\subsection{Scalability and Applicability of the Results}
Given the small scale of the laboratory setup and the use of a different method to measure the wave spectrum than what may be used in the field, some discussion of the scalability and applicability of the results reported herein is warranted. 

The absolute accuracy achieved herein with the PEDM is related to the scale of the setup, as well as the characteristics of the wave spectrum. The more pertinent metric is the fractional improvement in accuracy relative to the EDM, which is expected to be scalable to larger measurement setups and different techniques of measuring the wave spectrum. The relative improvement using the PEDM is related to the form of the current profile. In cases where the profile is approximately linear over the range of depths, limited improvement is expected since the approximation to which the EDM's mapping function was based is valid. In cases where the current profile has greater curvature near the surface, the PEDM is found to yield a greater fractional improvement in accuracy. The PEDM thus acts in a sense to improve the estimate to the current profile where possible, while performing similarly with the EDM otherwise. Note that the shape of the lab current profiles in the stagnation region, profiles a-c) in Figure~\ref{fig:upiv}, are representative of a scaled-down surface shear layer such as may be produced in the wind-swept ocean Ekman layer or in a river delta plume such as reported by \cite{kilcher10}. They differ in shape only by a constant subtraction of the deep-water velocity which corresponds to a constant offset in Doppler shifts. 

\begin{figure}[htb]
\centering
	\includegraphics{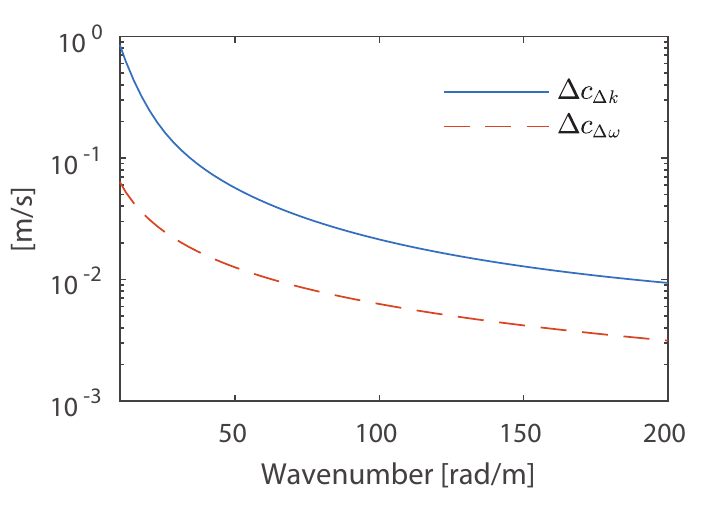}
	\caption{Doppler shift velocity bounds based on the pixel sizes in Fourier space, resulting from the spatial and temporal extent of the measurement domain.
	}
	\label{fig:vdelk}
\end{figure}

\rev{\subsubsection{\drev{Error sources}{Resolution}}}

The absolute accuracy of the Doppler shifts is fundamentally determined by, among other factors, the size of the measurement domain \rev{$L_x$} which sets the resolution in $k$-space, $\Delta k$\rev{$=2\pi/L_x$}. Herein, the extraction of the Doppler shift velocities was performed by evaluating the wave spectrum on a surface in spatiotemporal Fourier space with wavenumber $k$ kept constant, requiring interpolation between the available discrete values of \drv{$k$}{$\{k_x,k_y\}$}. \drv{Smaller values}{A smaller value} of $\Delta k$ reduces errors due to interpolation and also decreases the spectral leakage from neighboring wavenumber components. In an attempt to bound the \drev{errors}{uncertainties} in Doppler shifts caused by interpolation we define a velocity shift $\Delta c_{\Delta k}$ so that 

\begin{equation}
k \Delta c_{\Delta k}(k) = \frac{\mathrm{d} \omega_0(k)}{\mathrm{d} k}\Delta k.
\end{equation}
$\Delta c_{\Delta k}(k)$ is thus the depth-uniform current velocity that causes the linear dispersion surface to move by approximately one pixel in $k$-space for the relevant constant frequency $\omega_0(k)$. The values of $
\Delta c_{\Delta k}$ over the range of wavenumbers where Doppler shifts were extracted are shown in Figure~\ref{fig:vdelk} (see also Figure~\ref{fig:ctil_exp} for the Doppler shift wavenumber range). Another source of \drev{error}{uncertainty} in the Doppler shift involves the spread of the spectrum in frequency space, related to $\Delta\omega = 2\pi/T$\rev{, where $T$ is the total measurement period}. Again, we transform this quantity to a velocity:
\begin{equation}
\Delta c_{\Delta\omega}(k) = \Delta\omega/k,
\end{equation}
which is also shown in Figure~\ref{fig:vdelk}. Given the sizes of our measurement domain in space and time, $\Delta c_{\Delta k}$ is nearly an order of magnitude greater than $\Delta c_{\Delta \omega}$ over the range of relevant wavenumbers, indicating that resolution in $k$-space is the main contribution to \drev{errors}{uncertainties} in the Doppler shifts. Examining the figure gives an estimate to the upper bounds to the \drev{errors}{uncertainties} that can be expected in the Doppler shifts, and similarly the reconstructed profiles due to the finite spectral resolution. Comparing the values of $\Delta c_{\Delta k}$ to the values of $\Delta U_\mathrm{RMS}$ from the PEDM, it is evident that a great degree of sub-pixel resolution is achieved using the NSP and PEDM methods: $\Delta U_\mathrm{RMS}$ is less than the minimum value of $\Delta c_{\Delta k}$ for all current profiles, being orders of magnitude less than values of $\Delta c_{\Delta k}$ for the lower wavenumbers. 

We note that values of $\Delta c_{\Delta k}$ and $\Delta c_{\Delta \omega}$ for full-scale measurements in the ocean using for example X-band radar are typically within an order-of-magnitude of the values shown in Figure~\ref{fig:vdelk}, assuming \rev{spatial domain size} $L_x \sim 750$ m, $T \sim 10$ min, and wavenumbers in the range of $0.05-0.3$ $\rev{\rev{\mathrm{rad}}\cdot}\mathrm{m}^{-1}$ as is common \citep[e.g.][]{lund15}. Thus, though values of $\Delta U_\mathrm{RMS}$ from measurements in the ocean at large scales are expected to be larger than those reported here, it is not expected that the errors will increase by orders-of-magnitude.

%
%
%

\rev{\subsubsection{Scalability}}

Consider now how the small-scale experimental \drv{set-up}{setup} scales up to an oceanographic scale. First, it is obvious that the one effect which does not scale up, is that of surface tension, which is utterly negligible at the wavelengths measurable with e.g.\ X-band radar. In our experiment we do observe Bond number $\rho g \lambda^2/\sigma \lesssim \mathcal{O}(1)$ at the shortest wavelengths, yet the majority of our spectrum lies in the gravity wave regime, thus being physically directly comparable. This said, the PEDM method is not sensitive to whether or not the dispersion relation has capillary corrections at high $k$, and so the stringency of our testing is little altered by this. 

Assuming wavelengths to lie in the gravity wave regime, and assuming essentially infinite depth as is approximately true of our experiment, the system scales in the following way. Now only a single nondimensional group remains, a shear-Froude number based on three physical parameters: a typical wavelength of the spectrum, $g$, and a suitably defined depth-averaged shear. A \drev{natural alternative}{suitable definition} is
\begin{equation}
  \mathrm{Fr}_S(k)
  = \frac{1}{\sqrt{gk}}\int_{-\infty}^0\mathrm{d} z \mathbf{k}\cdot\mathbf{U}'(z)\mathrm{e}^{2kz} = \frac{\langle S\rangle_k}{\omega_0(k)},
\end{equation}
referred to as $\delta$ by \citet{ellingsen17}. $\langle S\rangle_k$ is the depth averaged shear along $\mathbf{k}$ suitably weighted for wave number $k$. 
Full similarity can be obtained if, by scaling up the velocity profile to oceanographic scale, the range of important $k$-values in the wave spectrum yields the same values of $\mathrm{Fr}_S$. Let's assume $U(z)$ is the lab current, and an oceanographic current of the same shape is $U_O(z) = u^* U(\delta z)$ with $\delta$ a small parameter describing the slower variation with depth and $u^*$ the fraction of the velocities at $z=0$. To probe the velocity profile into the depth in a similar manner as before, a lower wave number (i.e.\ longer wavelength) $k' = \delta k$ is required. On the whole we obtain $\mathrm{Fr}_S \to u^* \sqrt{\delta} \mathrm{Fr}_S$. In other words, similarity is in order if $u^*\sqrt{\delta}\sim\mathcal{O}(1)$.

Our most strongly sheared velocity profile, in Figure~\ref{fig:pedm_u}a), resembles in shape and magnitude a very strong oceanographic velocity profile, such as that can be found in the Columbia River delta \citep{kilcher10}, if we let $\delta = 1/500$ and $u^* = 12$, for example, resulting in $u^*\sqrt{\delta}\sim 0.54$ and shear-Froude numbers of the same order of magnitude. Wavelengths $500$ times those of the lab are reasonable for waves in the area, between $8$ and $80$ m for the wave numbers of Figure~\ref{fig:ctil_exp}. Hence we conclude that, while the strongest shear tested in the lab is a little stronger than can be expected of a particularly strong scaled-up equivalent, it is a satisfactory test of the PEDM theory in realistic settings. Given the ease of high quality flow measurements, scaled-down lab experiments thus offer an ideal test-bed for studies of ocean wave propagation on shear currents. 

We now comment on the range of depths at which the near-surface current profile is estimated. The depth range is determined directly by the range of mapped depths, and hence the range of wavenumbers in the measured spectrum. Though the choice of the mapping function is in a sense arbitrary, we argue the choice is reasonable based on intuition considering (\ref{eq:SJ}). At a depth $(2k)^{-1}$ the cumulative integral of the weighting function $2k\mathrm{e}^{2kz}$ is 0.63, i.e. a wave is influenced by roughly comparable amounts by currents at greater vs. shallower depths, indicating a reasonable choice of the depth assignment for most current profiles. Given the rapidly decreasing sensitivity of waves to currents at greater depths, the polynomial fits of the PEDM can be considered to be an expansion of the near-surface current profile in the top layer of the water column, valid over the depth range of the mapped Doppler shifts. As is well-known with polynomial fits, large errors can result with extrapolation for prediction of currents at greater depths.  In the laboratory experiments reported here, the depth range of the reconstructed flow is only a few centimeters, while in the ocean with wave spectra measured by X-band radar the depths may extend to tens of meters, given a roughly three orders of magnitude increase in the scale of the measured wavenumbers. 

\rev{\subsubsection{Wave spectrum measurement}}
For the laboratory results presented here, the wave spectrum was measured using a synthetic Schlieren method which measures directly the gradient components of the free surface, differing from methods that are practical for field measurements on a larger scale. However, for the purposes of inversion methods, all that is required is a signal that has the same periodicity in space and times as the wave spectrum. As mentioned in the introduction, various methods of measuring the wave spectrum in the radar and optical regime have already been used in reconstructing near surface currents. The choice of the wave spectral measurement method affects primarily the range of wavenumbers that are probed and is relatively inconsequential in terms of the inversion method process, affecting only the details of extraction of the Doppler shifts. A main difference between field measurement techniques such as X-band radar and the SS is that the mapping of free surface elevation to measured signal is, to a greater degree, nonlinear. The nonlinearities result in a signal at higher harmonics in the wave spectrum, yet the fundamental harmonic has the same periodicity in space and time as true wave component. Furthermore, the NSP method uses the signal at the second harmonic in determining the Doppler shifts. Thus, though the SS method employed in this work is impractical to be used in field measurements at larger scales, it can be viewed as an equivalent technique to those used in the field for the purposes here of fundamentally studying inversion methods.

\rv{\subsubsection{Applications}

The PEDM method may be applied to Doppler shifts extracted from wave spectra obtained by observation techniques readily available with today's technology, such as X-band radar or optical images of the ocean surface, as well as potential future methods for remotely sensing the directional wave spectrum. The Doppler shifts may be extracted by a number of means such as least squares techniques or the NSP method described and further developed herein.

As demonstrated in figure \ref{fig:pedm_u}, the PEDM offers greatest improvement in accuracy over the EDM in cases where the current profile has strong near-surface curvature within the range of mapped Doppler shifts. For the case of wave spectra measured by X-band radar where the mapped depths may typically be on the order of 2-10 m \citep[e.g.][]{lund15}, current profiles with strong curvature are expected to occur in times of high winds, and at specific locations such as river deltas with strong shear currents driven by density differences in the fluid \citep[e.g.][]{kilcher10}. Use of the PEDM to achieve a more accurate current depth-profile under such circumstances could result in improved characterization of submesoscale currents \citep{lund18},improved estimates of wave steepness for predicting breaking waves \citep{zippel17}, and improved mapping of shear currents for coastal engineering applications, for example. Under extreme sea states such as during hurricanes, improved accuracy in the reconstruction of remotely-sensed shear current profiles could allow for better prediction of wave and current forces on structures, where \citet{dalrymple73} has shown that currents even with velocities small compared to the wave orbital velocities can result in a notable increase in the forces on structures. In the latter case, however, strong wave nonlinearity and imaging difficulties may make remote sensing difficult in practice. 

For wave spectra measured using optical-based methods, the range of mapped depths is typically significantly shallower than for X-band radar data, in some cases resolving the top few centimeters of the water column where the current may have strong curvature even under moderate conditions \citep{laxague18}. The PEDM has the potential to improve the accuracy of the reconstruction in such cases, furthering applications such as studies of the air-sea interaction as well as the transport of contaminants near the ocean surface \citep{laxague18}.

In conditions where the current profile is approximately depth-uniform over the range of mapped depths, the PEDM is not expected to increase the accuracy of the reconstructed currents compared to the EDM or other existing methods which assume depth-uniform flow, yet figure \ref{fig:pedm_uy} demonstrates that the PEDM gives essentially identical results in such cases, eschewing the need to employ different methods in different conditions. By being simple to employ and performing equally well or better than current methods, we propose that the PEDM can replace competing inversion methods in current use in most situations. The exception we can imagine is situations where calculation cost is a very severe restriction.

\subsubsection{Limitations and challenges}
As with all inversion methods, the absolute accuracy of the PEDM is affected by the wave spectrum bandwidth in terms of wavenumber and angular spread. Reconstruction of the depth profile of the flow places more stringent demands on the wave spectrum having a broader range of wavenumbers and directions, when compared to methods aimed at estimating a single (depth-uniform) velocity vector, given the additional fitting parameters associated with the PEDM method: the PEDM involves $n_{\mathrm{max}} + 1$ polynomial coefficients for each horizontal dimension, whereas depth-uniform estimation requires only one. The need for a sufficient spectrum of waves to be present, however, is due to fundamental physics and will affect any method whereby currents are estimated from surface wave dispersion. If the currents have no surface imprint, clearly they simply cannot be inferred from surface measurement. Likewise, sufficient image quality is a fundamental requirement for all methods. 

In addition, under some circumstances such as extreme sea states nonlinear wave interactions become more prevalent, in which case analysis of the wave spectrum becomes more complicated due to the presence of bound waves. The same complication has also been observed for moderate wave slopes in a wind wave tank \citep{laxague17}. Analyzing the wave spectrum to extract the Doppler shifts corresponding to currents when nonlinear wave interactions are prevalent requires further study.

The PEDM method, like other similar methods which it aspires to replace, assumes horizontally homogeneous currents. When the horizontal variation is not slow compared to all relevant wavelengths, such as will often be the case particularly in coastal areas, more advanced methods will be required, beyond the current state-of-the-art. 

}

\section{Conclusions}

A new method for reconstructing near surface current profiles from measurements of the wave spectrum has been presented, demonstrated and carefully tested and compared to the state-of-the-art \rv{inversion method}.

The method is easy to implement. It takes the present state-of-the art technique of assigning effective depths to measured Doppler shift velocities (the effective depth method, EDM) as its starting point. A polynomial fit is made to the EDM profile from whose coefficients a new velocity profile estimate of polynomial form is created via a simple derived relation.
The resulting polynomial profile is an improved estimate to the true current profile compared to state-of-the-art methods such as the EDM as it does not make any \emph{a priori} assumptions on the general shape of the profile, and involves very little added complexity.


Our new polynomial effective depth method (PEDM) was tested on data obtained from a laboratory setup where background currents of different depth profiles could be created in a controlled manner and measured independently using particle image velocimetry which was used as ``truth'' measurements. The laboratory setup is an ideal test-bed for further studies regarding remote sensing of near-surface shear currents given the large degree to which the current profile and wave spectrum can be controlled and the straightforward scalability of the results up to oceanic scales. The PEDM offers a $>3\times$ improvement in accuracy relative to the EDM for profiles with strong near-surface curvature. For cases where the true current profile has approximately constant shear, the assumptions upon which the EDM is based are fulfilled, and the PEDM offers limited improvement in accuracy. The estimate produced is then similar to that of the EDM in accuracy and shape, demonstrating the robustness of the method.

A simple criterion was developed to determine optimal values for parameters involved in the polynomial fits to achieve the most probable current profile estimate. The criterion depends on the measured Doppler shift data only, and thus the PEDM involves no free parameters. A novel adaptation of the normalized scalar product method (NSP) was developed to extract Doppler shifts from wave spectra at multiple wavenumbers\drv{ using}{, including} the second harmonic \drv{to increase sensitivity to currents}{of the spectrum}.

The results indicate that the method can be applied to full scale field measurements to obtain higher accuracy in reconstructing near surface shear profiles from the wave spectrum, beneficial across a wide variety of oceanic applications.

\acknowledgments
S\AA E and A\AA\  were funded by the Research Council of Norway (FRINATEK), grant no.\ 249740. Data and scripts used for the analysis are available at: \rv{\url{https://doi.org/10.18710/8JBWCJ} (doi: 10.18710/8JBWCJ)}.


\end{document}